\documentclass[prb,preprint]{revtex4}
\usepackage{graphicx}
\begin{document}
\draft
\begin{abstract}
We show that the dynamics of an elastic solid embedded in a
Minkowski space consist of a set of coupled equations describing a
spin-$1/2$ field, $\Psi$, obeying Dirac's equation, a vector
potential, $A_\mu$, obeying Maxwell's equations and a metric,
$g_{\mu\nu}$, which satisfies the Einstein field equations. The
combined set of Dirac's, Maxwell's and the Einstein field equations
all emerge from a simple elastic model in which the field variables
$\Psi$, $A_\mu$ and $g_{\mu\nu}$ are each identified as derived
quantities from the field displacements of ordinary elasticity
theory.  By quantizing the elastic field displacements, a
quantization of all of the derived fields are obtained even though
they do not explicitly appear in the Lagrangian. We demonstrate the
approach in a three dimensional setting where explicit solutions of
the Dirac field in terms of fractional derivatives are obtained. A
higher dimensional version of the theory would provide an alternate
approach to theories of quantum gravity.
\end{abstract}
\pacs{PACS numbers: } \vspace{.5in}
\title{A New Approach to Quantum Gravity from a Model of an Elastic Solid}
\author{John M. Baker}
\affiliation{ 2221 Parnassus Ct; Hayward, CA 94542} \maketitle

\section{Introduction}In constructing a quantum field theory, the
usual prescription is to start with a known equation of motion, such
as Dirac's equation, and "invent" a suitable Lagrangian that
reproduces the equation of motion when Lagrange's equations are
applied. While this prescription has been successful it is not
unique.  In other words it is possible for two different
Lagrangian's to lead to the same equation of motion. For example,
section $12-2$ of Reference[$1$] provides a good example of two
different Lagrangian's that lead to the same equation of motion for
the density variations in an acoustic field.

In a theory of quantum gravity, this traditional approach would
involve using a Lagrangian with an appropriate set of terms such
that when Lagrange's equations are applied, the Einstein field
equations are reproduced. The Lagrangian obtained in this manner
explicitly contains the gravitational metric, as well as any other
field variables that are coupled to it. For instance an attempt at
merging gravity with QED would produce a Lagrangian that explicitly
includes the Dirac Spinor field $\Psi$, the electromagnetic vector
potential $A_\mu$ and the gravitational metric $g_{\mu\nu}$.

In this paper we demonstrate an alternate approach to quantum
gravity based on a model of an elastic solid.  In the this model,
the only field variables that appear in the Lagrangian are the field
displacements, $u_i$, that occur in elasticity theory. Using the
methods of fractional calculus, we will show that the equations of
motion of the
 system describe excitations that
can be identified as massless, non-interacting, spin-$1/2$ particles
obeying Dirac's equation. We then assume that one of our coordinates
is periodic and use a dimensional reduction technique to reduce the
dimensionality from three-dimensions to two-dimensions.

When terms beyond the linear approximation are included, this
dimensional reduction produces a new set of equations in which the
spin field, $\Psi$, is shown to interact via a vector potential
$A_\mu$ and a metric with field variables $g_{\mu\nu}$. The
compatibility equations of St. Venant are shown to reproduce
Maxwell's equation for $A_\mu$ and the Einstein field equations for
$g_{\mu\nu}$. We quantize the field displacements using standard
approaches and thereby produce a quantization of $\Psi$, $A_\mu$ and
$g_{\mu\nu}$ even though none of these quantities appears explicitly
in the Lagrangian.  We demonstrate the basic methods in a
three-dimensional setting where exact expressions for the Dirac
field can be obtained.

When quantized, this theory provides a low dimensional version of a
quantum description electrodynamics coupled to gravity.  If this
procedure could be extended to higher dimensions it would provide an
alternate approach to theories of quantum gravity.

\section{Elasticity Theory}
\label{sec:elasticity_theory}
 The theory of elasticity is usually
concerned with the infinitesimal deformations of an elastic
body\cite{ref:Love,ref:Sokolnikoff,ref:Landau_Lifshitz,ref:Green_Zerna,ref:Novozhilov}.
We assume that the material points of a body are continuous and can
be assigned a unique label $\vec{a}$. For a three-dimensional solid
each point of the body may be labeled with three coordinate numbers
$a^{i}$ with $i=1,2,3$.

If this three dimensional elastic body is placed in a large ambient
three dimensional space then the material coordinates $a^{i}$ can be
described by their positions in the 3-D fixed space coordinates
$x^{i}$ with $i=1,2,3$.  We imagine that the solid is free to
distort within the fixed ambient space described with coordinates
$x^i$.  In this description the material points $a^{i}(x^1,x^2,x^3)$
are functions of $\vec{x}$. A deformation of the elastic body
results in infinitesimal displacements of these material points. If
before deformation, a material point $a^0$ is located at fixed space
coordinates $x^{01},x^{02},x^{03}$ then after deformation it will be
located at some other coordinate $x^1,x^2,x^3$.  The deformation of
the medium is characterized at each point by the displacement vector
\[u^i=x^i-x^{0i}
\]
which measures the displacement of each point in the body after
deformation.  We will assume that our elastic solid is periodic in
the coordinate $a^3$ and at various points in this paper we will
Fourier transform the $a^3$ coordinate.

It is one of the aims of this paper to take this model of an elastic
medium and derive from it equations of motion that have the same
form as Dirac's equation. In doing so we have to distinguish between
the intrinsic coordinates of the medium which we will call
"internal" coordinates and the fixed space coordinates which
facilitates our derivation of the equations of motion. In the
undeformed state we may take the external coordinates to coincide
with the material coordinates $a^i= x^{0i}$. The approach that we
will use in this paper is to derive equations of motion using the
fixed space coordinates and then translate this to the internal
coordinates of our space.

\subsection{Strain Tensor}
\label{sec:strain_tensor} Let us assume that we have an elastic
solid embedded in a three-dimensional Minkowski space with metric
 \begin{equation}
 \label{eq:minkowski_metric}
 \eta_{ij}= \left(\begin{array}{ccc}
 -1 & 0 & 0\\
 0 & 1 & 0\\
 0 & 0 & 1
\end{array}
\right)
\end{equation}

We first consider the effect of a deformation on the measurement of
distance. After the elastic body is deformed, the distances between
its points changes as measured with the fixed space coordinates. If
two points which are very close together are separated by a radius
vector $dx^{0i}$ before deformation, these same two points are
separated by a vector $dx^i=dx^{0i}+du^i$ afterwards. The squared
distance between the points before deformation is then
$ds^2=-(dx^{01})^2+(dx^{02})^2+(dx^{03})^2$. Since these coincide
with the material points in the undeformed state, this can be
written
$ds^2=-(da^1)^2+(da^2)^2+(da^3)^2=\sum_{i,j}da^i\eta_{ij}da^j$. The
squared distance after deformation can be
written\cite{ref:Landau_Lifshitz}

\begin{eqnarray*}
ds'^2&=&\sum_{ij}\left(dx^i\right)\eta_{ij}\left(dx^j\right)\\
&=& \sum_{ij}\left(da^i+du^i\right)\eta_{ij}\left(da^j+du^j\right)\\
 &=&\sum_{ij}\left(da^i +
\sum_k\frac{\partial u^i}{\partial a^k}da^k\right)\eta_{ij}
\left(da^j + \sum_l\frac{\partial u^j}{\partial
a^l}da^l\right)\\
&=&\sum_{ij}\eta_{ij}da^i da^j + \sum_{ijk}\eta_{ij}\frac{\partial
u^i}{\partial a^k}da^k da^j+ \sum_{ijl}\eta_{ij}\frac{\partial
u^j}{\partial a^l}da^i da^l + \sum_{ijkl}\eta_{ij}\frac{\partial
u^i}{\partial a^k}\frac{\partial
u^j}{\partial a^l}da^k da^l\\
&=&\sum_{ij} \left(\eta_{ij}+\sum_k\left(\eta_{ik}\frac{\partial
u^k}{\partial a^j}+\eta_{jk}\frac{\partial u^k}{\partial
a^i}\right)+\sum_{kl}\eta_{kl}\frac{\partial u^k}{\partial
a^i}\frac{\partial u^l}{\partial a^j}\right) da^i da^j\\
&=&\sum_{ik}\left(\eta_{ij}+2\epsilon_{ij}\right)da^i da^j
\end{eqnarray*}
 where $\epsilon_{ij}$ is

\begin{equation}
\label{eq:strain_tensor}
\epsilon_{ij}=\frac{1}{2}\sum_k\left(\eta_{ik}\frac{\partial
u^k}{\partial a^j}+\eta_{jk}\frac{\partial u^k}{\partial
a^i}+\sum_{l}\eta_{kl}\frac{\partial u^k}{\partial
a^i}\frac{\partial u^l}{\partial a^j}\right).
\end{equation} and the presence of the matrix $\eta_{ij}$ simply
reflects the fact that we are assuming our solid is embedded in a
Minkowski space with a pseudo-Euclidean metric.


 The quantity $\epsilon_{ik}$ is known as the strain tensor. It is
fundamental in the theory of elasticity. In the above derivation,
the material or internal coordinates were treated as functions of
the fixed space coordinates.  As is well known in elasticity theory,
we could just as well treat the fixed space coordinates as functions
of the material coordinates.  In this case, the strain tensor has
the form\cite{ref:Sokolnikoff}
\begin{equation}
\label{eq:strain_tensor_alternate}
\epsilon_{ij}=\frac{1}{2}\sum_k\left(\eta_{ik}\frac{\partial
u^k}{\partial x^j}+\eta_{jk}\frac{\partial u^k}{\partial
x^i}-\sum_{l} \eta_{kl}\frac{\partial u^k}{\partial
x^i}\frac{\partial u^l}{\partial x^j}\right)
\end{equation}
These two different approaches to the strain tensor are known in
elasticity theory as the Lagrangian and Eulerian perspectives.  In
this work we will derive the equations of motion using the fixed
space coordinates which simplifies the derivation and we will
translate the result, when necessary, to the internal coordinates.

In most treatments of elasticity it is assumed that the
displacements $u^i$ as well as their derivatives are infinitesimal
so the last term in Equation~(\ref{eq:strain_tensor}) is dropped. In
this work, we will treat the strain components as small but finite.
We will then examine the structure of the equations of motion when
the higher order terms are treated as small perturbations on the
infinitesimal strain results.

\subsection{Metric Tensor} \label{sec:metric_tensor}
The quantity
\begin{eqnarray}
\label{eq:metric_tensor}
 g_{ij}&=&\eta_{ij}+\sum_k\left(\eta_{ik}\frac{\partial u^k}{\partial
a^j}+\eta_{jk}\frac{\partial u^k}{\partial
a^i}+\sum_{l}\eta_{kl}\frac{\partial u^k}{\partial
a^i}\frac{\partial u^l}{\partial a^j}\right)\\
&=&\eta_{ij}+2\epsilon_{ij}\nonumber
\end{eqnarray}
 is the metric for our system and
determines the distance between any two points. One interesting
aspect of the elasticity theory approach is that it provides a
natural metric on the system in terms of the strain components
expressed entirely in terms of the internal coordinates of the
elastic body. This means that at any point in space the distance
measurement can be made without reference to the fixed space
coordinates.  In other words if you were an ant living in this
elastic medium, Equation~(\ref{eq:metric_tensor}) would be the
metric that you would use.

Even though the metric in Equation~(\ref{eq:metric_tensor}) does not
have the Euclidean form, the space in which we are working is still
intrinsically flat.  The metric that we derived is due simply to a
coordinate transformation and so cannot describe the curved space of
general relativity. That this metric is simply the result of a
coordinate transformation from the Minkowski metric can be seen by
writing the metric in the form\cite{ref:Millman_Parker}
\[
 g_{\mu\nu}=
 \left( \begin{array}{lll}{\displaystyle
\frac{\partial x^1}{\partial a^1}}& {\displaystyle\frac{\partial
x^2}{\partial
a^1}} & {\displaystyle\frac{\partial x^3}{\partial a^1}}\\[15pt]
 {\displaystyle\frac{\partial x^1}{\partial a^2}}& {\displaystyle\frac{\partial x^2}{\partial a^2}} &
{\displaystyle\frac{\partial x^3}{\partial a^2}}\\[15pt]
{\displaystyle\frac{\partial x^1}{\partial a^3}}&
{\displaystyle\frac{\partial x^2}{\partial a^3}} &
{\displaystyle\frac{\partial x^3}{\partial a^3}}
\end{array}
 \right)
 \left(\begin{array}{rrr}
 {\displaystyle -1}& {\displaystyle 0} & {\displaystyle 0}\\[15pt]
 {\displaystyle 0}& {\displaystyle 1}& {\displaystyle 0}\\[15 pt]
 {\displaystyle 0}& {\displaystyle 0} & {\displaystyle 1}
 \end{array}
 \right)
 \left(\begin{array}{lll}
{\displaystyle \frac{\partial x^1}{\partial a^1}}&
{\displaystyle\frac{\partial x^1}{\partial
a^2}} & {\displaystyle \frac{\partial x^1}{\partial a^3}}\\[15pt]
 {\displaystyle\frac{\partial x^2}{\partial a^1}}& {\displaystyle\frac{\partial x^2}{\partial a^2}} &
{\displaystyle\frac{\partial x^2}{\partial a^3}}\\[15pt]
{\displaystyle\frac{\partial x^3}{\partial a^1}}&
{\displaystyle\frac{\partial x^3}{\partial a^2}} &
{\displaystyle\frac{\partial x^3}{\partial a^3}}
\end{array}
\right)
\]
\[
=J^T\eta J
\]
where
\[
\frac{\partial x^\mu}{\partial a^\nu}=\delta_{\mu\nu}+\frac{\partial
u^\mu}{\partial a^\nu}.
\]
and $J$ is the Jacobian of the transformation. Later in section
\ref{sec:internal_coordinates}, however we will use a dimensional
reduction technique borrowed from Kaluza-Klein theories to reduce
the three-dimensional flat space to a two-dimensional curved space.
We will show that the metric for the Fourier modes of this two
dimensional system is not a simple coordinate transformation.

The inverse metric, which is written with upper indices as $g^{ik}$,
can be obtained by explicitly inverting
Equation~(\ref{eq:metric_tensor}) or we can write
$(g^{ik})=(J^{-1})\eta^{-1} (J^{-1})^T$ where
\begin{equation}
J^{-1}=\left(\begin{array}{lll} {\displaystyle\frac{\partial
a^1}{\partial x^1}}& {\displaystyle\frac{\partial a^1}{\partial
x^2}} & {\displaystyle\frac{\partial a^1}{\partial x^3}}\\[15pt]
 {\displaystyle\frac{\partial a^2}{\partial x^1}}& {\displaystyle\frac{\partial a^2}{\partial x^2}} &
{\displaystyle\frac{\partial a^2}{\partial x^3}}\\[15pt]
{\displaystyle\frac{\partial a^3}{\partial x^1}}&
{\displaystyle\frac{\partial a^3}{\partial x^2}} &
{\displaystyle\frac{\partial a^3}{\partial x^3}}
\end{array}
\right)
\end{equation}
 This yields for the inverse metric
 \begin{eqnarray}
 \label{eq:inverse_metric}
 g^{\mu\nu}&=&\eta_{\mu\nu}+\sum_\alpha\left(-\eta_{\nu\alpha}\frac{\partial
u^\mu}{\partial x^\alpha}-\eta_{\mu\alpha}\frac{\partial
u^\nu}{\partial x^\alpha}+\sum_{\alpha \beta}\eta_{\alpha\beta}
\frac{\partial u^\mu}{\partial x^\alpha}\frac{\partial
u^\nu}{\partial x^\beta}\right)
 \end{eqnarray}
Equation~(\ref{eq:inverse_metric}) shows that we can write the
inverse matrix directly in terms of derivatives of $u_i$ with
respect to the fixed space coordinates. This form of the inverse
metric will be useful in later sections.

\subsubsection{Internal vs. External Coordinates and Summation
Convention} \label{internal_external} The change in form of the
metric between that given in Equation~(\ref{eq:minkowski_metric})
and Equation~(\ref{eq:metric_tensor}) is due simply to a change in
coordinates between the fixed space coordinates and the material
coordinates.  In this regard the transformation is similar to
changing from Cartesian to spherical coordinates.  This change is
useful because it allows us to derive equations in the fixed space
coordinates where the calculations are simplified, and then when
necessary we can switch to the internal coordinates using
$u^i=x^i-a^i$.

We would like to be able to use the notation that a raised index on
a variable indicates a contraction with the metric tensor and that a
raised index and a lower index with the same label implies a
summation (ie the Einstein summation convention). We have to be
careful, however, to point out which set of coordinates, and hence
which metric we are using, so we will be explicit in each section as
to which coordinate system the raised indices refer to. For instance
we can write Equation~(\ref{eq:strain_tensor}) more compactly as
\begin{equation}
\label{eq:strain_tensor_raised}
\epsilon_{ij}=\frac{1}{2}\left(\frac{\partial u_i}{\partial
a^j}+\frac{\partial u_j}{\partial a^i}+\frac{\partial u_l}{\partial
a^i}\frac{\partial u^l}{\partial a^j}\right).
\end{equation}
where $u_l u^l= \sum_k\eta_{lk} u_l u^k$ so that upper/lower indices
indicate contraction with the fixed space metric,
Equation~(\ref{eq:minkowski_metric}).

\subsection{St Venant's Equations of Compatibility}
In Section~(\ref{sec:EOM}) we will derive the equations of motion of
the elastic solid using the Lagrangian formalism.  There are,
however, additional constraints that an elastic solid must also
satisfy. These constraints are called the St Venant equations of
compatibility in classical elasticity theory\cite{ref:Sokolnikoff}.
The usual description of these compatibility equations is that they
are integrability conditions or, a restriction on the strain
components $\epsilon_{ij}$ such that they can be considered partial
derivatives of a function $u$ as displayed in
Equation~(\ref{eq:strain_tensor}). In other words if $\epsilon_{ij}$
is a function that is composed of the partial derivatives of $u$
then it has to satisfy certain conditions and these are the
compatibility equations.

However, from a geometric standpoint these equations are simply a
re-statement of the fact that space is flat\cite{ref:Novozhilov}. In
other words the compatibility equations are equivalent
to\cite{ref:Novozhilov}
\begin{equation}
\label{eq:riemann}
R_{\alpha\beta\mu\nu}\equiv\frac{1}{2}\left(\frac{\partial^2
g_{\alpha\nu}}{\partial a^\beta a^\mu}-\frac{\partial^2
g_{\alpha\mu}}{\partial a^\beta a^\nu}+\frac{\partial^2
g_{\beta\mu}}{\partial a^\alpha a^\nu} - \frac{\partial^2
g_{\beta\nu}}{\partial a^\alpha a^\mu}\right)+
\Gamma^\rho_{~\beta\mu}\Gamma_{\rho,\alpha\nu}
-\Gamma^\rho_{~\alpha\mu}\Gamma_{\rho,\beta\nu}=0,
\end{equation}
 where $R_{\alpha\beta\mu\nu}$ is the Riemann Curvature tensor
 and $\Gamma^\rho_{~\alpha\beta}$ are the Christoffel
symbols given by\cite{ref:Schutz}
\[
\Gamma^\rho_{~\alpha\beta}=g^{\lambda\rho}\left(\frac{\partial
g_{\lambda\alpha}}{\partial x^\beta}+ \frac{\partial
g_{\lambda\beta}}{\partial x^\alpha}-\frac{\partial
g_{\alpha\beta}}{\partial x^\lambda}\right).
\]
In the above equations an upper/lower index implies a contraction
with the metric in the internal coordinates.

 One of the interesting
aspects of an elastic solid is that this setting gives you for
"free" an explicit expression for the metric,
Equation~(\ref{eq:metric_tensor}), and a statement about the
curvature of space, Equation~(\ref{eq:riemann}). We get these
equations even though, as we will see shortly, the metric is not a
dynamical variable appearing in the Lagrangian.

\subsection{Equation of Motion}
\label{sec:EOM}In the following we will use the notation
\[
u_{\mu\nu}=\frac{\partial u_\mu}{\partial x^\nu}
\]
 and therefore the strain tensor is
 \begin{equation}
 \label{eq:strain_tensor_restated}
 \epsilon_{\mu\nu}=\frac{1}{2}\left(u_{\mu\nu}+u_{\nu\mu}+ u_{l\mu}u^l_\nu\right).
 \end{equation} and all contractions are with the fixed space
 metric,
Equation~(\ref{eq:minkowski_metric}).

 We
work in the fixed space coordinates and take the negative of the
strain energy as the lagrangian density of our system. This approach
leads to the usual equations of equilibrium in elasticity
theory\cite{ref:Love,ref:Novozhilov}. The strain energy is quadratic
in the strain tensor $\epsilon^{\mu\nu}$ and therefore the
Lagrangian can be written
\[
L=-\sum_{\mu \nu\alpha\rho} C^{\mu \nu\alpha\rho}\,
\epsilon_{\mu\nu} \epsilon_{\alpha\rho}
\]

The quantities $C^{\mu \nu\alpha\rho}$ are known as the elastic
stiffness constants of the material\cite{ref:Sokolnikoff}.  For an
isotropic space most of the coefficients are zero and in fact there
are only two independent elastic constants in a three-dimensional
isotropic space. The lagrangian density then reduces to
\begin{equation}
\label{eq:lagrangian_3D}
 L=(\lambda +
2\mu)\left[\epsilon_{11}^2+\epsilon_{22}^2+\epsilon_{33}^2\right] +
2 \lambda \left[ \epsilon_{22}\epsilon_{33}-\epsilon_{11}
\epsilon_{22}- \epsilon_{11} \epsilon_{33}\right] + 4\mu \left[
\epsilon_{23}^2-\epsilon_{12}^2 - \epsilon_{13}^2 \right]
\end{equation}
where $\lambda$ and $\mu$ are known as Lam\'e
constants\cite{ref:Sokolnikoff}.

We first derive the equations of motion of the system in the
approximation where the strain components $u_{ij}$ are
infinitesimal.  In the infinitesimal strain approximation, the
quadratic terms in Equation~(\ref{eq:strain_tensor_restated}) are
dropped giving
\[
 \epsilon_{\mu\nu}=\frac{1}{2}\left(u_{\mu\nu}+u_{\nu\mu}\right).
\]

The usual Lagrange equations,
\[
\sum_\nu\frac{d}{dx^\nu}\left(\frac{\partial L}{\partial u_{\rho
\nu}}\right) - \frac{\partial L}{\partial u_\rho}=0,
\]
apply with each component of the displacement vector, $u^\rho$,
treated as an independent field variable.

Using the above form of the Lagrangian one can write
\[
\frac{\partial L}{\partial u_{\rho
\nu}}=2\lambda(\sigma\eta^{\rho\nu})+4\mu \epsilon^{\rho\nu}
\]
where the divergence of the displacement field is
$\sigma\equiv(-u_{11}+u_{22}+u_{33})$.  In classical elasticity
theory $\sigma$ is known as the dilatation and physically represents
the fractional change in density of a medium due to a deformation.

 We now have three field equations (one for each value of $\rho$),
\begin{equation}
\label{eq:field_eq} \frac{\partial}{\partial x^\nu}\frac{\partial
L}{\partial u_{\rho \nu}}=(2\lambda+2\mu)\frac{\partial}{\partial
x_\rho}\sigma +2\mu \nabla^2(u^\rho)=0
\end{equation}
where $\nabla^2=-\frac{\partial^2}{\partial
x_1^2}+\frac{\partial^2}{\partial x_2^2}+\frac{\partial^2}{\partial
x_3^2}$.  Applying the operator $\partial/\partial x^\rho$ to
Equation~(\ref{eq:field_eq}) yields the wave equation
\begin{equation}
\label{eq:wave_equation} \left(-\frac{\partial^2}{\partial
x_1^2}+\frac{\partial^2}{\partial x_2^2}+\frac{\partial^2}{\partial
x_3^2}\right)\sigma=0
\end{equation}

Equation~(\ref{eq:wave_equation}) shows that the classical
dilatation in the medium obeys the wave equation. In
Section~(\ref{sec:Dirac_EOM}) we will demonstrate a new method for
reducing the wave equation (\ref{eq:wave_equation}) to Dirac's
equation and compare this method to the traditional Dirac reduction.
But first we turn our attention to quantizing the field
displacements $u_i$ in this elastic model.

\subsection{Quantization} \label{sec:quantization}

The Lagrangian density of the system is given by
Equation~(\ref{eq:lagrangian_3D}). The coordinate $x_1$ plays the
role of time in this three-dimensional space so the canonical
momenta associated with the field variable $u_\rho$ are given by
\[
P_\rho=\frac{\partial L}{\partial u_{\rho1}}
\]
which gives

\begin{eqnarray}
\label{eq:canonical_momenta}
P_1&=&-2\lambda(\epsilon_{11}-\epsilon_{22}-\epsilon_{33})-4\mu\epsilon_{11}\\
P_2&=&4\mu\epsilon_{12}\nonumber\\
P_3&=&4\mu\epsilon_{13}\nonumber.
\end{eqnarray}

The Hamiltonian density is defined as
\begin{eqnarray*}
\mathcal{H}&=&\sum_\rho P_\rho u^\rho_1-L\\
&=&-P_1u_{11}+P_2u_{21}+P_3u_{31}-L
\end{eqnarray*}
 and the total Hamiltonian is the integral over all space of the Hamiltonian density
\[
H=\int d^3x \mathcal{H}
\]
 Inverting Equation~(\ref{eq:canonical_momenta}) allows us to
 replace the variables $u_{\rho 1}$ with
$P_\rho$ in the result.  This gives

\begin{eqnarray}
\label{eq:hamiltonian}
 \mathcal{H}&=&\frac{3 P_1^2}{4 (\lambda +2 \mu
)}+\frac{P_2^2+P_3^2}{4 \mu }-\frac{\lambda
\left(u_{22}+u_{33}\right)
   P_1}{\lambda +2 \mu }-P_2
   u_{12}-P_3 u_{13}\nonumber\\
   &&\hspace{.5in}+\mu  \left(u_{23}+u_{32}\right){}^2-\frac{\lambda ^2
   \left(u_{22}+u_{33}\right){}^2}{\lambda +2 \mu }+2 \lambda  u_{22} u_{33}\nonumber\\
    &&\hspace{.5in}+(\lambda +2 \mu )
   \left(u_{22}^2+u_{33}^2\right)
\end{eqnarray}
 We now Fourier transform the field variables making the assumption
that one of our coordinates, $a_3$ is compact with the topology of a
circle. Therefore, when we Fourier transform the field variables,
the $q_3$ component is associated with a discrete spectrum while the
other two coordinates are continuous.  Writing out the coordinate
dependencies explicitly we have,
\[
P_\rho(x_1,x_2,x_3)=\sum_{q_3}\int dq_1 \int dq_2
P_{\rho,\vec{q}}e^{\imath\vec{q}\cdot \vec{x}}
\]
\[
u_\rho(x_1,x_2,x_3)=\sum_{q_3}\int dq_1 \int dq_2
u_{\rho,\vec{q}}e^{\imath\vec{q}\cdot \vec{x}}.
\] The Fourier transform
results in terms in the Hamiltonian that mix field variables
associated with $q$ and $-q$. For instance, the contribution to the
total Hamiltonian from the $P_1^2$ term in
Equation~(\ref{eq:hamiltonian}) becomes
\begin{eqnarray*}
\int d^3x  P_1^2&=&\int d^3x\left(\sum_{q_3}\int dq_1 \int dq_2
\sum_{q'_3}\int dq'_1 \int dq'_2
P_{1,\vec{q}}P_{1,\vec{q'}}e^{\imath
(\vec{q}+\vec{q'})\cdot\vec{x}}\right)\\
&=&\sum_{q_3}\int dq_1 \int dq_2 P_{1,\vec{q}}P_{1,-\vec{q}}
\end{eqnarray*}

The total Hamiltonian then becomes
\[
H=\sum_q H_q
\]
where $H_q$ is written symmetrically in $q$ and $-q$ as
\begin{eqnarray}
\label{eq:hamiltonian_FT} H_q&=&-\frac{q_2^2 u_{2,-q} u_{2,q}
\lambda ^2}{\lambda +2 \mu }-\frac{q_2 q_3 u_{2,q} u_{3,-q}
   \lambda ^2}{\lambda +2 \mu }-\frac{q_2 q_3 u_{2,-q} u_{3,q} \lambda ^2}{\lambda +2 \mu
   }-\frac{q_3^2 u_{3,-q} u_{3,q} \lambda ^2}{\lambda +2 \mu }\\
   &&+\frac{i q_2 P_{1,q} u_{2,-q}
   \lambda }{2 (\lambda +2 \mu )}-\frac{i q_2 P_{1,-q} u_{2,q} \lambda }{2 (\lambda +2 \mu
   )}+q_2^2 u_{2,-q} u_{2,q} \lambda +\frac{i q_3 P_{1,q} u_{3,-q} \lambda }{2 (\lambda +2
   \mu )}\nonumber \\
   &&+q_2 q_3 u_{2,q} u_{3,-q} \lambda-\frac{i q_3 P_{1,-q} u_{3,q} \lambda }{2
   (\lambda +2 \mu )}+q_2 q_3 u_{2,-q} u_{3,q} \lambda +q_3^2 u_{3,-q} u_{3,q}
   \lambda\nonumber\\
   &&+\frac{3 P_{1,-q} P_{1,q}}{4 (\lambda +2 \mu )}+\frac{P_{2,-q} P_{2,q}}{4 \mu
   }+\frac{P_{3,-q} P_{3,q}}{4 \mu }+\frac{1}{2} i q_2 P_{2,q} u_{1,-q}\\
   &&+\frac{1}{2} i q_3
   P_{3,q} u_{1,-q}-\frac{1}{2} i q_2 P_{2,-q} u_{1,q}-\frac{1}{2} i q_3 P_{3,-q} u_{1,q}+2
   \mu  q_2^2 u_{2,-q} u_{2,q}\nonumber\\
   &&+\mu  q_3^2 u_{2,-q} u_{2,q}+\mu  q_2 q_3 u_{2,q}
   u_{3,-q}+\mu  q_2 q_3 u_{2,-q} u_{3,q}+\mu  q_2^2 u_{3,-q} u_{3,q}+2 \mu  q_3^2 u_{3,-q}
   u_{3,q}
\end{eqnarray}
Since terms in the Hamiltonian with different values of $q$ are not
mixed, the function $H_q$ in Equation~(\ref{eq:hamiltonian_FT}) can
be solved independently for each $q$.  $H_q$ is a bilinear function
in the variables $u_{i,\pm q}$ and $P_{i,\pm q}$ and can be
diagonalized exactly, using the methods of
Biougliobov\cite{ref:Kittel,ref:Tsallis,ref:Tikochinsky}.

\subsubsection{Exact Diagonalization} \label{sec:diagonalization} The
methods used in diagonalizing the Hamiltonian in
Equation~(\ref{eq:hamiltonian_FT}) are summarized in the
references\cite{ref:Tsallis,ref:Tikochinsky}. The idea is to rewrite
the Hamiltonian in terms of a set of creation and annihilation
operators, $b_{i,q}$ and $b_{i,q}^\dagger$ such that the Hamiltonian
has the form
\[
H_q=\sum_{i} \omega_{i,q}b^\dagger_{i,q} b_{i,q}
\]
and the operators satisfy the commutation relations
\[
[b_{i,q},b^\dagger_{j,q'}]=i\delta_{i,j}\delta_{q,q'}.
\]

The details of this procedure are included in
Appendix~(\ref{app:quantization}). Of particular interest are the
energy eigenvalues of the modes. There are three distinct positive
energies for the states $b_i$. They are
\begin{eqnarray}
\label{eq:eigenenergies}
E_{1,q}&=&\frac{1}{4} \sqrt{q_2^2+q_3^2}\\
E_{2,q}&=&\frac{1}{4} \sqrt{1-\frac{4 \mu  (\lambda +\mu
)}{\sqrt{\mu  (\lambda
   +\mu ) (\lambda +2 \mu )^2}}} \sqrt{q_2^2+q_3^2}\\
E_{3,q}&=&\frac{1}{4} \sqrt{1+\frac{4 \mu  (\lambda +\mu
)}{\sqrt{\mu  (\lambda
   +\mu ) (\lambda +2 \mu )^2}}} \sqrt{q_2^2+q_3^2}
\end{eqnarray}

With the operators $b_i$ calculated, the field variables $u_i$ can
be written as a linear combination of creation and annihilation
operators as
\begin{equation}
\label{eq:u_intermsof_b}
 u_i=\sum_{q=1}^3\sum_{i=1}^6\left(c_{ij} b_i
e^{i\vec{q}\cdot\vec{x}}+ c'_{ij} b^\dagger_i
e^{-i\vec{q}\cdot\vec{x}}\right)
\end{equation}
where the $c_{ij}$ are coefficients given in the appendix and $q_1$
is the energy of a given mode. These eigenstates are the linear
approximation obtained from keeping the lowest terms in the strain
components $u_{ij}$ in Equation~(\ref{eq:lagrangian_3D}).  The
higher order terms that were left out of the Lagrangian can be
incorporated by treating them as perturbations. In other words we
can use standard perturbation theory to find new strain components
that are nonlinear in the creation and annihilation operators. These
field components can then serve as the basis for a theory of finite
strain as we do in the next section.

Strictly speaking the field displacements as expressed in
Equation~(\ref{eq:u_intermsof_b}) are not energy eigenstates since
the $b_i$ have different energies.  We will mainly be concerned
however, with a low energy approximation of the spectrum of this
elastic solid.  For positive values of $\lambda$ and $\mu$ and
$\lambda
>\left(6+4 \sqrt{3}\right) \mu$, the energies
$E_{3,q}$ can be arbitrarily small compared to $E_{1,q}$ and
$E_{2,q}$.   For instance with $\lambda=13\mu$ and $\mu=.1$, the
energies $E_{1,q}$ and $E_{2,q}$ are more than $20$ times greater
than $E_{3,q}$. This suggests that in a low energy theory only the
excitation corresponding to energies $E_{3,q}$ will be present for
suitably defined Lame constants. We will not investigate the
mechanical properties of such a solid but merely point out that in
such a theory at low energies, both the field displacements $u_i$
and the dilatation, $\nabla\cdot\vec{u}$, are energy eigenstates. We
also note that each of the energies is proportional to
$\sqrt{q_2^2+q_3^2}$ with $q_3$ taking on discrete values. So one
would expect in the lowest energy approximation that only the modes
with $q_3=0$ will be present and at slightly higher energies the
mode with $q_3=1$ will be present. This low energy approximation
will be exploited in later sections.

One of the things that we have gained from this formalism is the
ability to calculate any quantity, that depends on the field
displacements, quantum mechanically.  For instance we can now
calculate the metric given in Equation~(\ref{eq:inverse_metric})
using the form of the field decomposition given in
Equation~(\ref{eq:u_intermsof_b}) even though the metric itself is
not a dynamical variable appearing in the Lagrangian.


In the finite strain theory treated in
Section~(\ref{sec:finite_strain}) we will need to take Fourier
transforms of field variables in the internal coordinates rather
than the fixed space coordinates.  Since we will be keeping the
nonlinear terms in all of our equations, then for consistency we
assume that the field variables in Equation~(\ref{eq:u_intermsof_b})
have been properly treated to the same order in perturbation theory.
We will not explicitly calculate the field variables in perturbation
theory rather we will focus on the form of the field equations when
terms beyond the linear approximation are kept.

We will now give a new derivation of Dirac's equation as the
equation of motion of the elastic solid.

\section{Derivation of Dirac's Equation of Motion}
\label{sec:Dirac_EOM}

\subsection{Cartan's Spinors} \label{sec:Cartan}
 The concept of Spinors was introduced by Eli
Cartan in 1913\cite{ref:Cartan}. In Cartan's original formulation
spinors were motivated by studying isotropic vectors which are
vectors of zero length. In three dimensional Minkowski space the
equation of an isotropic vector is
\begin{equation}
\label{eq:isotropic_vector}
 -x_1^2 + x_2^2 + x_3^2=0
\end{equation}
for generally complex quantities $x_i$.  A closed form solution to
this equation is realized as
\begin{equation}
\label{eq:Cartan_spinor_solution}
\begin{array}{lccr}
{\displaystyle x_1=\xi_0^2+\xi_1^2,\ } & {\displaystyle
x_2=\xi_0^2-\xi_1^2,}&\ \mathrm{and}\ & {\displaystyle x_3
=-2\xi_0\xi_1}
\end{array}
\end{equation}
 where the two quantities $\xi_i$ are then
\[
\begin{array}{lcr}
{\displaystyle\xi_0=\pm\sqrt{\frac{x_1+ x_2}{2}}}& \ \mathrm{and} \
& {\displaystyle \xi_1=\pm\sqrt{\frac{x_1- x_2}{2}}}
\end{array}.
\]
The two component object $\xi=(\xi_0,\xi_1)$ has the rotational
properties of a spinor\cite{ref:Cartan} and any equation of the form
(\ref{eq:isotropic_vector}) has a spinor solution.

In the following we use the notation
$\partial_\mu\equiv\partial/\partial x^\mu$ and the wave equation is
written
\[\left(-\partial_1^2+\partial_2^2+\partial_3^2\right)\phi=0.
\]
This equation can be viewed as an isotropic vector in the following
way. The components of the vector are the partial derivative
operators $\partial/\partial x^\mu$ acting on the quantity $\phi$.
As long as the partial derivatives are restricted to acting on the
scalar field $\phi$ it has a spinor solution given by
\begin{equation}
\label{eq:spinor0}
 \hat{\xi}_0^2=\frac{1}{2}\left(\frac{\partial}{\partial
x^1}+\frac{\partial}{\partial x^2}\right)
\end{equation}
and
\begin{equation}
\label{eq:spinor1}
\hat{\xi}_1^2=\frac{1}{2}\left(\frac{\partial}{\partial
x^1}-\frac{\partial}{\partial x^2}\right)
\end{equation} where the "hat" notation indicates that the quantities
$\hat{\xi}$ are operators. Let us now introduce the variables
\[
\begin{array}{lcr}
 {\displaystyle z_0=x_1+x_2}& \ \mathrm{and}\ & {\displaystyle
 z_1=x_1-x_2}
\end{array}
\]

Equations~(\ref{eq:spinor0}) and (\ref{eq:spinor1}) are now
\[
\hat{\xi}_0^2=\frac{\partial}{\partial z^0}
\]
and
\[
\hat{\xi}_1^2=\frac{\partial}{\partial z^1}.
\]
These are equations of fractional derivatives of order $1/2$ (also
called semiderivatives) denoted $\hat{\xi}_0=D^{1/2}_{z^0}$ and
$\hat{\xi}_1=D^{1/2}_{z^1}$. Fractional derivatives have the
property that\cite{ref:Miller_Ross}
\[
D^{1/2}_{z}D^{1/2}_{z}=\frac{\partial}{\partial z}
\]
 and various methods exist for writing closed form solutions for these
 operators\cite{ref:Miller_Ross,ref:Oldham_Spanier}.
The exact form for these fractional derivatives however, is not
important here.  The important thing to note is that a solution to
the wave equation can be written in terms of spinors which are
fractional derivatives.

One of the interesting properties of suitably defined fractional
derivatives (for instance the Weyl fractional derivative) that will
be exploited in later sections is their action on the exponential
function. While the derivative of an exponential is given by
\[
\frac{\partial}{\partial x}e^{\alpha x}=\alpha e^{\alpha x}
\]
the semiderivative of the exponential function is given by
\begin{equation}
\label{eq:FT_semiderivative} D^{1/2}_{x}e^{\alpha
x}=\sqrt{\alpha}e^{\alpha x}
\end{equation}
This will prove useful later when we Fourier transform the equations
of motion.

\subsection{Matrix Form}
It can be readily verified that our spinors satisfy the following
equations
\begin{eqnarray*}
\left[\hat{\xi}_0 \frac{\partial}{\partial x^3}+ \hat{\xi}_1
\left(\frac{\partial}{\partial x^1}+\frac{\partial}{\partial
x^2}\right)\right]\phi=0\\
 \left[\hat{\xi}_0\left(\frac{\partial}{\partial x^2} -
\frac{\partial}{\partial
x^1}\right)-\hat{\xi}_1\frac{\partial}{\partial x^3}\right]\phi=0
\end{eqnarray*}
and in matrix form
\begin{equation}
\label{eq:dirac_matrix}
 \left(
  \begin{array}{lr}
{\displaystyle\frac{\partial}{\partial x^3}} &
{\displaystyle\frac{\partial}{\partial x^1}+\frac{\partial}{\partial
 x^2}} \\[15pt]
{\displaystyle\frac{\partial}{\partial x^2}-\frac{\partial}{\partial
 x^1}} & {\displaystyle -\frac{\partial}{\partial x^3}}
  \end{array}
   \right)
 \left(\begin {array}{c}
 {\displaystyle\hat{\xi}_0 }\\[20pt]
  {\displaystyle\hat{\xi}_1}
  \end{array}
  \right)  \phi=0
\end{equation}

The matrix
\[
 X=\left( \begin{array}{lr}
{\displaystyle\frac{\partial}{\partial x^3}} &
{\displaystyle\frac{\partial}{\partial x^1}+\frac{\partial}{\partial
 x^2}} \\[15pt]
{\displaystyle\frac{\partial}{\partial x^2}-\frac{\partial}{\partial
 x^1}} & {\displaystyle -\frac{\partial}{\partial x^3}}
  \end{array}
 \right)
 \]
 is equal to the dot product of the vector $\partial_\mu\equiv\partial/\partial x^\mu$ with the pauli spin matrices
 \[
 X=\frac{\partial}{\partial x^1}\gamma^1 + \frac{\partial}{\partial
 x^2}\gamma^2 + \frac{\partial}{\partial x^3}\gamma^3
 \]
 where
 \[
 \begin{array}{ccc}
 \gamma^1=\left(\begin{array}{rr}
 0 & 1\\
 -1 & 0
 \end{array}
 \right),&
\gamma^2=\left(\begin{array}{ll}
 0 & 1\\
 1 & 0
 \end{array}
  \right),&
 \gamma^3=\left(\begin{array}{lr}
 1 & 0\\
 0 & -1
 \end{array}
 \right)
\end{array}
 \] are proportional to the Pauli matrices and satisfy the
 anticommutation relations
 \begin{equation}
\label{eq:anti_comm} \{\gamma^{\mu},\gamma^{\nu}\}=2I \eta^{\mu\nu}.
\end{equation} where $I$ is the identity matrix.

Equation~(\ref{eq:dirac_matrix}) can be written
\begin{equation}
\label{eq:dirac_unstrained}
\sum_{\mu=1}^3\partial_\mu\gamma^\mu\hat{\xi}\phi=0.
\end{equation}
This equation has the form of Dirac's equation in three-dimensions
for a noninteracting, massless, spin-$1/2$ field, $\hat{\xi}\phi$.

\subsection{Relation to the Dirac Decomposition}
The fact that the wave equation and Dirac's equation are related is
not new.  However the decomposition used here is not the same as
that used by Dirac. The usual method of connecting the second order
wave equation to the first order Dirac equation is to operate on
Equation~(\ref{eq:dirac_unstrained}) from the left with
$\sum_{\nu=1}^3\gamma^{\nu}\partial_\nu$ giving

\begin{eqnarray}
\label{eq:dirac_to_wave}
0&=&\sum_{\mu,\nu=1}^3\gamma^{\nu}\gamma^{\mu}\partial_\nu \partial_\mu\Psi(x)\nonumber\\
&=&\sum_{\mu,\nu=1}^3\frac{1}{2}\left(\gamma^{\nu}\gamma^{\mu}+\gamma^{\mu}\gamma^{\nu}\right)
\partial_\nu\partial_\mu\Psi(x) \nonumber \\
&=&\left(-\partial_1^2+\partial_2^2+\partial_3^2\right)\Psi(x)
\end{eqnarray}
where $\Psi=(\alpha_1,\alpha_2)$ is a two component spinor and
Equation~(\ref{eq:anti_comm}) has been used in the last step.

This shows that Dirac's equation does in fact imply the wave
equation. The important thing to note about
Equation~(\ref{eq:dirac_to_wave}) however, is that the
three-dimensional Dirac's equation implies not one wave equation but
two in the sense that each component of the spinor $\Psi$ satisfies
this equation. Explicitly stated, Equation~(\ref{eq:dirac_to_wave})
reads
\[
\left(\begin{array}{cc}
 -\partial_1^2+\partial_2^2+\partial_3^2 & 0\\
 0 & -\partial_1^2+\partial_2^2+\partial_3^2
 \end{array}
 \right)\left(\begin{array}{c}
 \alpha_1\\
 \alpha_2\end{array}\right) =0
 \] for the independent scalars $\alpha_1, \alpha_2$.

Conversely, if one starts with the wave equation and tries to
recover Dirac's equation, it is necessary to start with two
independent scalars each independently satisfying the wave equation.
In other words, using the usual methods, it is not possible to take
a single scalar field that satisfies the wave equation and recover
Dirac's equation for a two component spinor.

What has been demonstrated in the preceding sections is that
starting with only one scalar quantity satisfying the wave equation,
Dirac's equation for a two component spinor may be derived.
Furthermore any medium (such as an elastic solid) that has a single
scalar that satisfies the wave equation must have a spinor that
satisfies Dirac's equation and such a derivation necessitates the
use of fractional derivatives.

\section{Dimensional Reduction in Infinitesimal strain}
\label{sec:FT_infinitesimal} In this section we take a closer look
at the equation of motion, Equation~(\ref{eq:dirac_unstrained}), and
the spinor (represented as a fractional derivative) when the field
displacements are Fourier transformed.  In the infinitesimal theory
of elasticity, all terms in $u_{ij}$ beyond the linear term are
dropped.  In the infinitesimal theory therefore, no distinction is
made between transforming the coordinates $a_i$ and $x_i$.  Later
when we assume small but finite strain components, we will need to
distinguish these coordinates.

The Dirac field in Equation~(\ref{eq:dirac_unstrained}) is given
explicitly by
\begin{equation}
\label{eq:spinor_Before Eq}
\hat{\xi}\phi(x_1,x_2,x_3)=\left(\begin{array}{l}
D^{1/2}_{z_0} \\
D^{1/2}_{z_1}
\end{array}\right)\phi(x_1,x_2,x_3)
\end{equation}
with $z_0=x_1+x_2$ and $z_1=x_1-x_2$.  Because the semiderivatives
$D^{1/2}_{z_0,z_1}$ are independent of $x_3$ we can bring $e^{\imath
x_3}$ through the operator $\hat{\xi}$ when we Fourier transform
$\phi$.

 Transforming the
dilatation first in the periodic coordinate $q_3$,
Equation~(\ref{eq:dirac_unstrained}) becomes

\begin{eqnarray}
\label{eq:Deq_infinitesimal}
0&=&\sum_{\mu=1}^3\gamma^\mu\partial_\mu\hat{\xi}\phi\nonumber\\
&=&\left(\sum_{\mu=1}^2\gamma^\mu\partial_\mu+\gamma^3\partial_3\right)\hat{\xi}\phi\nonumber\\
&=&\sum_q e^{\imath q_3
x_3}\left(\sum_{\mu=1}^2\gamma^\mu\partial_\mu+\gamma^3\imath
q_3\right)\hat{\xi}\phi_{q_3}\nonumber\\
&=&\sum_q e^{\imath q_3
x_3}\left(\sum_{\mu=1}^2\gamma^3\gamma^\mu\partial_\mu+\imath
q_3\right)\hat{\xi}\phi_{q_3}
\end{eqnarray}
where we used $\phi=\sum_{q_3}\phi_{q_3}e^{\imath q_3 x_3}$, and $
\gamma^3\gamma^3=1$.

Equation~(\ref{eq:Deq_infinitesimal}) is equal to zero only if the
coefficients of $e^{\imath q_3 x_3}$ are zero for each value of
$q_3$ giving
\begin{equation}
\label{Deq_infinitesimal_2D} \left(\sum_{\mu=1}^2\imath
\gamma^{\prime\mu}\partial_\mu- q_3\right)\hat{\xi}\phi_{q_3}=0
\end{equation}
where $\gamma^{\prime\mu}=\gamma^3\gamma^\mu$ and satisfies the
conditions for a two dimensional metric
$\left\{\gamma^{\prime\mu},\gamma^{\prime\nu}\right\}=2I\eta^{\prime\mu\nu}$
with
\begin{equation}
 \label{eq:minkowski_2D_metric}
 \eta^{\prime\mu\nu}= \left(\begin{array}{cc}
 1 & 0\\
 0 & -1\\
\end{array}
\right)
\end{equation}
Equation~(\ref{Deq_infinitesimal_2D}) shows that the fourier modes
of the elastic solid obey a two dimensional version of Dirac's
equation for spin-$1/2$ particles with a mass $q_3$.  The two
continuous variables left in the problem are $x_1$ and $x_2$ with
$x_1$ playing the role of time.

Let us examine the form of the spinor further by Fourier
transforming the two continuous coordinates $x_1$ and $x_2$. Using
Equation~(\ref{eq:FT_semiderivative}) we further transform the
spinor as
\begin{eqnarray}
\label{eq:spinor_Before Eq}
\hat{\xi}\phi_{q_3}(x_1,x_2)&=&\left(\begin{array}{l}
D^{1/2}_{z_0} \\
D^{1/2}_{z_1}
\end{array}\right)\phi_{q_3}(x_1,x_2)\nonumber\\
&=&\int dq_1 dq_2 \left(\begin{array}{l}
D^{1/2}_{z_0}e^{\imath (q_1 x_1+q_2x_2)}\\
D^{1/2}_{z_1}e^{\imath (q_1 x_1+q_2x_2)}
\end{array}\right)\phi_{q_1,q_2,q_3}\nonumber\\
&=&\int dq_1 dq_2 \left(\begin{array}{l}
D^{1/2}_{z_0}e^{\frac{1}{2}\imath z_0 (q_1+q_2)+\frac{1}{2}\imath z_1 (q_1-q_2)}\nonumber\\
D^{1/2}_{z_1}e^{\frac{1}{2}\imath z_0(q_1+q_2)+\frac{1}{2}\imath z_1
(q_1-q_2)}
\end{array}\right)\phi_{q_1,q_2,q_3}\\
&=&\int dq_1 dq_2 \,e^{\imath (q_1
x_1+q_2x_2)}\left(\begin{array}{l}
\sqrt{q_1+q_2}\\
\sqrt{q_1-q_2}
\end{array}\right)\sqrt{\imath}\phi_{q_1,q_2,q_3}
\end{eqnarray}

When the fields $b_i$ are viewed as a time dependent quantity (ie
the Heisenberg picture in quantum mechanics), the wavevector $q_1$
is equal to the energy of the $q^{th}$
mode\cite{ref:Sakurai,ref:Peskin}. This allows us to write the
column vector in Equation~(\ref{eq:spinor_Before Eq}) as
\[
u(q)=\left(\begin{array}{l}
\sqrt{E+q_2}\\
\sqrt{E-q_2}
\end{array}\right)\\
\]

Compare this formula to the expression for the four dimensional
Dirac spinor in a definite state of helicty\cite{ref:Peskin}
\[
u(p)=\left(\begin{array}{l}
\sqrt{E+P}\left(\begin{array}{c}0\\1\end{array}\right)\\
\sqrt{E-P}\left(\begin{array}{c}0\\1\end{array}\right)
\end{array}\right)\\
\]

The dimensional
 reduction in the infinitesimal theory of elasticity has produced a
 two dimensional version of Dirac's equation for a particle with a
 bare mass $q_3$ and a spinor that has a consistent form to the
 known $4$ dimensional version.

 Incidentally, with the identification $q_3$ as a mass term, the
 energy eigenstates given in Equation~(\ref{eq:eigenenergies}) are
 seen to have the relativistic form $E_q\sim\sqrt{q_2^2+m^2}$.

 The only problem with the interpretation as relativistic particles
 is the question of quantum statistics.  Relativistic spin-$1/2$
 particles obey anticommutation relations while the boson operators
 $b_i$ that we have defined obey commutation relations. This
 situation could be "rescued" by defining new operators
 \[
c_k=\theta_k b_k
 \]
 where the $\theta_k$ are complex Grassman numbers.  Grassman numbers satisfy
\[
\begin{array}{lr}
 \theta_k\theta_{k'}=-
 \theta_{k'}\theta_k,&\hspace{.1in}\left(\theta_k\right)^2=0
 \end{array}
 \]
this would imply
\begin{eqnarray*}
\left\{c_k,c_{k'}\right\}&=&\left\{\theta_k b_k,\theta_{k'}
b_{k'}\right\}\\
&=&\theta_k\theta_{k'}b_k b_{k'}+\theta_{k'}\theta_{k}b_{k'} b_{k}\\
&=&\theta_k\theta_{k'}b_k b_{k'}-\theta_{k}\theta_{k'}b_{k'} b_{k}\\
&=&\theta_k\theta_{k'}\left[b_k, b_{k'}\right]\\
 &=&0
\end{eqnarray*}

Similarly we have,
\[
\begin{array}{lr}
\displaystyle
 \left\{c^\dagger_k,c^\dagger_{k'}\right\}=0, ~\mathrm{ and} &\hspace{.2in}
 \left\{c,c^\dagger_{k'}\right\}=\imath\theta_k\theta^\star_{k'}\delta_{k,k'}
 \end{array}
\]

With this definition, the fields $c_k$ satisfy appropriate quantum
statistics and still obeys Dirac's equation.  We do not wish to
dwell on this admittedly ad-hoc procedure for getting fermion
statistics into this theory. Rather we wish to focus on the form of
the equations of motion that are derived and demonstrate that the
all ingredients for a theory of quantum gravity are present in this
model.

 In this section, we have demonstrated that the equation of motion of this
 model of an elastic
 solid, in the infinitesimal strain approximation
 has as its equation of motion a two dimensional version of Dirac's
 equation for spin-$1/2$ particles.  However, these spin-$1/2$ particles described in
 Equation~(\ref{Deq_infinitesimal_2D}) do not interact with each
 other or the outside world.  In the next section we will treat the
 dimensional reduction problem again in the context of the finite
 theory of strain.  When terms beyond the linear approximation are
 kept, we show that the equations obtained describe particles that
 interact gravitationally and electromagnetically.

\section{Finite Strain}
\label{sec:finite_strain}

\subsection{Internal Coordinates}
\label{sec:internal_coordinates} In this section we will need to
Fourier transform our field variables in $a_3$ and therefore need to
translate the equations of motion from the fixed space coordinates
to the internal coordinates. For clarity and to adopt a more
consistent convention, in the remainder of this text we change
notation and write the internal coordinates not as $a^i$ but as
$x'^i$ and the fixed space coordinates will continue to be unprimed
and denoted $x^i$. Now using $u^i=x^i-x'^i$ we can write

\begin{eqnarray}
\label{eq:coordinate_change} \frac{\partial}{\partial
x^i}&=&\sum_j\frac{\partial x'^j}{\partial
x^i}\frac{\partial}{\partial x'^j} \nonumber \\
&=& \sum_j\left(\frac{\partial
x^j}{\partial x^i}-\frac{\partial u^j}{\partial x^i}\right)\frac{\partial}{\partial x'^j}\nonumber\\
&=& \sum_j\left(\delta_{ij}-\frac{\partial u^j}{\partial
x^i}\right)\frac{\partial}{\partial x'^j}\nonumber\\
&=& \frac{\partial}{\partial x'^i}-\sum_j\frac{\partial
u^j}{\partial x^i}\frac{\partial}{\partial x'^j}
\end{eqnarray}

Equation~(\ref{eq:coordinate_change}) relates derivatives in the
fixed space coordinates $x^i$ to derivatives in the material
coordinates $x'^i$. We can now re-write the three-dimensional
Dirac's equation as
\begin{eqnarray}
\label{eq:dirac_before_FT}
 \sum_{\mu=1}^3\gamma^\mu
\partial_\mu\Psi&=&\sum_{\mu=1}^3\gamma^\mu\left(\partial_\mu'-\sum_\nu\frac{\partial
u^\nu}{\partial x^\mu} \partial_\nu'\right)\Psi\nonumber\\
&=&\sum_{\mu=1}^3\gamma^{\prime\mu}\partial'_\mu\Psi=0
\end{eqnarray}
where $\Psi\equiv\hat{\xi}\phi$, $\partial'_\mu=\partial/\partial
x'_\mu$ and $\gamma^{\prime\mu}$ is given by
\begin{equation}
\label{eq:modified_gamma_matrices}
\gamma^{\prime\mu}=\gamma^\mu-\sum_{\alpha=1}^3
u^\mu_{~\alpha}\gamma^\alpha.
\end{equation} The $\gamma^{\prime\mu}$ are simply
the gamma matrices expressed in the primed coordinate system. The
anticommutator of these matrices is
\begin{eqnarray*}
\{\gamma^{\prime\mu},\gamma^{\prime\nu}\}&=&
\{\gamma^\mu-\sum_\alpha u^\mu_{~\alpha}\gamma^\alpha,\gamma^\nu-\sum_\beta u^\nu_{~\beta}\gamma^\beta\}\\
&=&\{\gamma^\mu,\gamma^\nu\}-\sum_\beta
u^\nu_{~\beta}\{\gamma^\mu,\gamma^\beta\} - \sum_\alpha
u^\mu_{~\alpha}\{\gamma^\alpha,\gamma^\nu\}+
\sum_{\alpha\beta}u^\mu_{~\alpha}u^\nu_{~\beta}\{
\gamma^\alpha,\gamma^\beta\}\\
&=&2I\left(\eta^{\mu\nu}-\sum_\beta u^\nu_{~\beta}\eta^{\mu\beta}-
\sum_\alpha u^\mu_{~\alpha}\eta^{\alpha\nu}+ \sum_\alpha\sum_\beta
u^\mu_{~\alpha}u^\nu_{~_\beta} \eta^{\alpha\beta}\right)\\
\end{eqnarray*}
Comparison with Equation~(\ref{eq:inverse_metric}) shows that
\begin{equation}
\label{eq:gamma_anticommutator}
\{\gamma^{\prime\mu},\gamma^{\prime\nu}\}=2Ig^{\mu\nu}
\end{equation}

 These gamma matrices have the form of the
usual dirac's matrices in a curved space\cite{ref:Brill_Wheeler}.
  To further develop the form of
Equation~(\ref{eq:dirac_before_FT}) we have to transform the spinor
properties of $\xi$.  Much like a normal vector, the components of a
spinor are altered under a change of coordinates and as currently
written $\xi$ is a spinor with respect to the $x_i$ coordinates not
the $x'_i$ coordinates. To transform its spinor properties we assume
(similar to Brill and Wheler\cite{ref:Brill_Wheeler}) a real
similarity transformation and write $\Psi=S\Psi^\prime$ where $S$ is
a transformation that takes the spinor in $x_\mu$ to a spinor in
$x'_\mu$.

We then have
\[
\partial'_\mu\Psi=(\partial'_\mu S)\Psi^\prime+S\partial'_\mu\Psi^\prime.
\]
Equation~(\ref{eq:dirac_before_FT}) then becomes
\[
\begin{array}{lcl}
 0&=&\gamma^{\prime\mu}
[S\partial'_\mu\Psi^\prime+(\partial'_\mu S)\Psi^\prime]\\
 \mbox{}
&=&\gamma^{\prime\mu}
S[\partial'_\mu\Psi^\prime+S^{-1}(\partial'_\mu
S)\Psi^\prime]\\
\mbox{} &=&S^{-1}\gamma^{\prime\mu}
S[\partial'_\mu\Psi^\prime+S^{-1}(\partial'_\mu S)\Psi^\prime]
\end{array}
\]
Using $(\partial'_\mu S^{-1}) S=-S^{-1}(\partial'_\mu S)$. This can
finally be written
\begin{equation}
\label{eq:dirac_curved_space} \tilde{\gamma}^\mu
[\partial'_\mu-\Gamma_\mu]\Psi^\prime=0
\end{equation}
where $\Gamma_\mu=(\partial'_\mu S^{-1})S$ and
$\tilde{\gamma}^\mu=S^{-1}\gamma^{\prime\mu} S$. The new gamma
matrices $\tilde{\gamma}^\mu$ still satisfy the appropriate
anticommutation condition
\begin{equation}
\label{eq:tilde_gamma}
\left\{\tilde{\gamma}^\mu,\tilde{\gamma}^\nu\right\}=2Ig^{\mu\nu}
\end{equation}

What we have done in the above manipulations is re-write
Equation~(\ref{eq:dirac_unstrained}) completely in terms of the
internal coordinates of the elastic solid.
Equation~(\ref{eq:dirac_curved_space}) has the same physical content
as Equation~(\ref{eq:dirac_unstrained}), it is just expressed in
different coordinates.

Notice however that Equation~(\ref{eq:dirac_curved_space}) now
superficially has the form of the Einstein-Dirac equation in
three-dimensions for a massless, noninteracting, spin-$1/2$
particle\cite{ref:Brill_Wheeler}. The quantity
$\partial'_\mu-\Gamma_\mu$ is the covariant derivative for an object
with spin. In order to make this identification, the field
$\Gamma_\mu$ must satisfy the additional
equation\cite{ref:Brill_Wheeler,ref:Brill_Cohen}
\[
\frac{\partial {\tilde{\gamma}^\mu}}{\partial
x^\nu}+\tilde{\gamma}^\beta\Gamma^{\prime\mu}_{\beta\nu}-
\Gamma_\nu\tilde{\gamma}^\mu+\tilde{\gamma}^\mu\Gamma_\nu=0
\]
where $\Gamma^{\prime\mu}_{\beta\nu}$ is the usual Christoffel
symbol. That this equation holds is shown in
Appendix~(\ref{app:Auxiliary_Equation}).

%
%
Just as was done in the infinitesimal strain approach we will use a
dimensional reduction to transform
Equation~(\ref{eq:dirac_curved_space}).  In this case however, the
gamma matrices couple the metric into the problem and this metric
will also need to be reduced from three to two dimensions.

\section{Dimensional Reduction in Finite Strain}
\label{sec:internal_coordinates}

The dimensional reduction method that we use is borrowed from Kaluza
Klein theory.  It consists of two parts.  First we need to remove
the dependence of the field variables on the coordinate
$x_3^\prime$.  This is accomplished by Fourier transforming the
field variables as before.  The second part consists of reducing the
$3\times 3$ matrix, $g_{\mu\nu}$, in three dimensions to a $2\times
2$ matrix suitable for a two dimensional space.  It is this
reduction of the metric that will introduce the electromagnetic
vector potential into our equations.

Let us denote the $3\times 3$, three dimensional metric as
$\tilde{g}_{\mu\nu}$ and the $2\times 2$, two dimensional metric as
$g_{\mu\nu}$. We now write the Kaluza Klein
ansatz\cite{ref:Wesson,ref:Liu_Wesson}
\begin{equation}
\label{eq:KK_metric}
 \tilde{g}_{\alpha\beta}=\left(\begin{array}{cc}
{\displaystyle
g_{\alpha\beta}-\Phi^2 A_\alpha}\hspace{.1in}& {\displaystyle-\Phi^2 A_\alpha}\\[15pt]
 {\displaystyle -\Phi^2A_\beta}
 & {\displaystyle-\Phi^2}
\end{array}
 \right)
 \hspace{.3in}
 \tilde{g}^{\alpha\beta}=\left(\begin{array}{cc}
 {\displaystyle g^{\alpha\beta}}& {\displaystyle -A^\alpha}\\[15pt]
 {\displaystyle -A^\beta}& \hspace{.1in}{\displaystyle \left(-\Phi^{-2}+A^\mu A_\mu\right)}
 \end{array}
 \right)
 \end{equation}
where the vector $A_\mu$ is the electromagnetic vector potential,
$\Phi$ is a scalar and the upper/lower indices indicate a
contraction with $g_{\mu\nu}$.

 With this ansatz for the metric, it can be shown
that the flat space condition $R_{\alpha\beta}=0$ in three
dimensions, reduces to the Einstein Field equations in curved
two-dimensional space and Maxwell's equations for the field $A_\mu$.
The details of this derivation are given by Liu and
Wesson\cite{ref:Liu_Wesson} where the reduction is performed in a
five dimensional setting.  This is a standard dimensional reduction
used in Kaluza Klein theories and has been shown to produce a
consistent treatment of four-dimensional gravity and
electromagnetism from five-dimensional flat space. Therefore, we
will not repeat this derivation here only noting that there is
nothing in the derivation that is particular to five
 dimensions.  The exact same treatment works in a dimensional
 reduction from three to two dimensions so that we will freely
 quote the results of that work which gives the following equations
 for the metric and the electromagnetic vector potential
%
\begin{eqnarray*}
F^\lambda_{~\alpha;\lambda}&=&-3\Phi^{-1}\Phi^\lambda F_{\lambda\alpha}\\
R_{\alpha\beta}&=&-\frac{1}{2}\Phi^2F_{\lambda\beta}F_\beta^\lambda+\Phi^{-1}\Phi_{\alpha;\beta}
\end{eqnarray*}
where $F_{\alpha\beta}\equiv A_{\beta;\alpha}-A_{\alpha;\beta}$ and
we have used the notation where a comma indicates an ordinary
derivative and a semicolon indicates covariant differentiation.

The second of these equations, may be written in the more
traditional form\cite{ref:Liu_Wesson}
\begin{equation}
\label{eq:field equations} G^{\alpha\beta}\equiv
R^{\alpha\beta}-\frac{1}{2}g^{\alpha\beta}R=
8\pi\left(T_{em}^{\alpha\beta}+T_s^{\alpha\beta}\right)
\end{equation}
where the quantities $T_{em}^{\alpha\beta}$ and $T_s^{\alpha\beta}$
are effective energy momentum tensors given by
\begin{eqnarray*}
T_{em}^{\alpha\beta}&=&-\frac{1}{2}\Phi^2 \left(F^\alpha_\lambda
F^{\beta\lambda}-\frac{1}{4}g^{\alpha\beta}
F^{\mu\nu}F_{\mu\nu}\right)\\
T_s^{\alpha\beta}&=&\Phi^{-1}\left(\Phi^{\alpha;\beta}-g^{\alpha\beta}\Phi^\mu_{;\mu}\right)
\end{eqnarray*}

So what we see is that by using the dimensional reduction technique
of Kaluza Klein theory we automatically obtained the Einstein field
equations and Maxwell's equations.  What we now show is that the
same dimensional reduction technique not only produces these
equations but also changes the non-interacting Einstein-Dirac
equation in three dimensions, into an interacting theory for a
massive spin-$1/2$ particle in two dimensions.

\subsection{Dimensional Reduction of the Dirac Equation}
We begin by rewriting Equation~(\ref{eq:dirac_curved_space}) as
\begin{eqnarray}
\label{Dirac_eq2}
 \sum_{\mu=1}^2\tilde{\gamma}^\mu
[\partial'_\mu-\Gamma_\mu]\Psi^\prime+\tilde{\gamma}^3
[\partial'_3-\Gamma_3]\Psi^\prime=0
\end{eqnarray}
We now rewrite the matrix $\tilde{\gamma}^3$.  From
Equations~(\ref{eq:KK_metric}) and (\ref{eq:tilde_gamma}) we can
write
\[
\tilde{\gamma}^3\tilde{\gamma}^3=g^{33}=A^\mu A_\mu-\Phi^{-2}
\]
This allows us to write
\begin{equation}
\label{eq:new_gamma3}
\tilde{\gamma^3}=\sum_{\alpha=1}^2\tilde{\gamma}^\alpha
A_\alpha+\gamma_\bot\Phi^{-1}
\end{equation}
where the matrix $\gamma_\bot$ satisfies
\[
\begin{array}{ccc}
\left\{\tilde{\gamma}^1,\gamma_\bot\right\}=0,&
\left\{\tilde{\gamma}^2,\gamma_\bot\right\}=0,& \gamma_\bot^2=1
\end{array}
\]
An explicit expression for $\gamma_\bot$ is given in
Appendix~(\ref{app:gamma_perp}). It can be seen that this form for
$\tilde{\gamma}^3$ correctly gives
$\tilde{\gamma}^3\tilde{\gamma}^3=g^{33}$.  We can now write
Equation~(\ref{Dirac_eq2}) as
\begin{eqnarray}
\label{Dirac_eq3} 0&=& \sum_{\mu=1}^2\tilde{\gamma}^\mu
[\partial'_\mu-\Gamma_\mu]\Psi^\prime+(\sum_{\alpha=1}^2\tilde{\gamma}^\alpha
A_\alpha+\gamma_\bot\Phi^{-1}) [\partial'_3-\Gamma_3]\Psi^\prime\nonumber\\
&=&\sum_{\mu=1}^2\tilde{\gamma}^\mu
[\partial'_\mu+A_\mu(\partial'_3-\Gamma_3)-\Gamma_\mu]\Psi^\prime+\gamma_\bot\Phi^{-1}
[\partial'_3-\Gamma_3]\Psi^\prime\nonumber\\
&=&\sum_{\mu=1}^2\gamma''^\mu
[\partial'_\mu+A_\mu(\partial'_3-\Gamma_3)-\Gamma_\mu]\Psi^\prime+\Phi^{-1}
[\partial'_3-\Gamma_3]\Psi^\prime
\end{eqnarray}
where $\gamma''^\mu=\gamma_\bot \tilde{\gamma}^\mu$. This equation
has the same form as the classical Einstein-Dirac-Maxwell
equation\cite{ref:Brill_Wheeler,ref:Brill_Cohen,ref:Finster,ref:Smoller_Finster,ref:Smoller_Finster2}.
The metric enters in the equation through the $\gamma''^\mu$
matrices which satisfy $
\left\{\gamma''^\mu,\gamma''^\nu\right\}=2Ig^{\mu\nu}$ and the
electromagnetic vector potential enters in the same way as the
minimal coupling prescription
$\partial_\mu\rightarrow\partial_\mu+\imath e/c A_\mu$.
Additionally, a mass term has been created in
Equation~(\ref{Dirac_eq3}) with $m=\Phi^{-1}
[\partial'_3-\Gamma_3]$.


The field variables in Equation~(\ref{Dirac_eq3}) still are
dependent on $x'_3$.  To complete the derivation we need to remove
this dependence by Fourier Transforming our field variables.  We
first write the spinor field
\[
\Psi=\sum_q \Psi_q e^{\imath qx'_3}
\]
this gives
\[
\sum_q e^{\imath qx'_3}\left\{\sum_{\mu=1}^2\gamma''^\mu
[\partial'_\mu+A_\mu(\imath
q-\Gamma_3)-\Gamma_\mu]\Psi^\prime+\Phi^{-1} [\imath
q-\Gamma_3]\Psi^\prime\right\}=0
\]
Now we write
\[
\begin{array}{cccc}
A_\mu=\sum_k A_{\mu,k}e^{\imath k x'_3}, &
\Gamma_\mu=\sum_k\Gamma_{\mu,k}e^{\imath k x'_3},&
\gamma''^\mu=\sum_{k'}\gamma''^\mu_{k'}\,e^{\imath k' x'_3},&
\Phi^{-1}=\sum_{k'}\Phi^{-1}_{k'}e^{\imath k' x'_3}
\end{array}
\]
Which gives
\begin{eqnarray*}
&&\sum_{q,k,k',k''}e^{\imath
x'_3(q+k+k'+k'')}\left\{\sum_{\mu=1}^2\gamma''^\mu_{k'}
[\partial'_\mu\delta_{k,0}\delta_{k'',0}+A_{\mu,k}(\imath
q\delta_{k'',0}-\Gamma_{3,k''})-\Gamma_{\mu,k}\delta_{k'',0}]\Psi_q^{\prime}\right.\nonumber\\
 & & \mbox{}\hspace{2.5in}\left.+\delta_{k'',0}\Phi_{k'}^{-1} [\imath
q\delta_{k,0}-\Gamma_{3,k}]\Psi_q^\prime\right\}=0
\end{eqnarray*}
 This equation is true only if the coefficients of the exponential are independently true.
Let $m=q+k+k'+k''$ then
\begin{eqnarray}
\label{eq:many_modes}
&&\sum_{k,k',k''}\left\{\sum_{\mu=1}^2\gamma''^\mu_{k'}
[\partial'_\mu\delta_{k,0}\delta_{k'',0}+A_{\mu,k}(\imath
(m-k-k'-k'')\delta_{k'',0}-\Gamma_{3,k''})-\Gamma_{\mu,k}\delta_{k'',0}]\Psi_{m-k-k'-k''}^{\prime}\right.\nonumber\\
 & & \mbox{}\hspace{1in}\left.+\delta_{k'',0}\Phi_{k'}^{-1} [\imath
(m-k-k'-k'')\delta_{k,0}-\Gamma_{3,k}]\Psi_{m-k-k'-k''}^\prime\right\}=0.
\end{eqnarray}
This is a series of equations, one for each distinct value of $m$.
The dynamics contained in Equation~(\ref{eq:many_modes}) describes
an infinite series of spin particles $\Psi_{m-k-k'-k''}$ interacting
via a infinite series of vector potentials $A_{\mu,k}$. Up until now
we have kept all terms in this series of equations.  We will now
examine Equation~(\ref{eq:many_modes}) in the low energy
approximation.

\subsection{Spectrum of Lowest modes}
\label{sec:lowest_modes} Each of the quantities in
Equation~(\ref{eq:many_modes}) is a function of the field
displacements $u_i$.  We showed in previous sections that the energy
of the Fourier modes of field displacements increases with $q_3$. We
therefore expect that in a system where the energy is arbitrarily
low, not all of the modes in Equation~(\ref{eq:many_modes}) will be
excited. At the lowest energies only the mode $\vec{u}_{q=0}$ will
be excited, as the energy of the system increases modes
$\vec{u}_{q=\pm1}$ becomes excited and so on.  As this energy is
increased Equation~(\ref{eq:many_modes}) rapidly becomes more
complex but at sufficiently low energies its form is quite simple.

We illustrate this by considering a theory in which only the lowest
modes are present. Strictly speaking, our Hamiltonian formalism in
Section~(\ref{sec:quantization}) produced energy eigenstates for
modes $\vec{u}_q$ in a basis $e^{\imath qx}$ while we want to make a
statement about the energy of the modes,$\vec{u}_{q'}$ in the basis
$e^{\imath q'x'}$, with $x'^i=x^i-u^i$. In
Appendix~(\ref{app:Fcomponents}) we show that the two are related
and that if the only mode present in the system is $q=0$ in the
first basis then, the only mode present is $q'=0$ in the second.

Let us now consider the case where there is insufficient energy to
excite the mode $ \vec{u}_{q=\pm1}$ and only $\vec{u}_{q=0}$ is
excited. This would represent the lowest energy possible in our
system. In this case the wavevectors in
Equation~(\ref{eq:many_modes}) must equal to $m=k=k'=k''=0$ and
Equation~(\ref{eq:many_modes}) reduces to

\begin{equation}
\label{eq:single_mode}\sum_{\mu=1}^2\gamma''^\mu_{0}
[\partial'_\mu-A_{\mu,0}\Gamma_{3,0}-\Gamma_{\mu,0}]\Psi_{0}^{\prime}
-\Phi_{0}^{-1} \Gamma_{3,0}\Psi_0^\prime=0.
\end{equation}

From Equation~(\ref{eq:inverse_metric}) we also have
\[
\begin{array}{lr}
\displaystyle
\tilde{g}^{\mu\nu}=\eta^{\mu\nu}+\sum_{\alpha=1}^2\left(-\eta^{\nu\alpha}\frac{\partial
u^\mu_{0}}{\partial x^\alpha}-\eta^{\mu\alpha}\frac{\partial
u^\nu_{0}}{\partial x^\alpha}+\sum_{\beta=1}^2\eta^{\alpha\beta}
\frac{\partial u^\mu_{0}}{\partial x^\alpha}\frac{\partial
u^\nu_{0}}{\partial x^\beta}\right),&\hspace{.1in}(\mu,\nu=1,2,3)
\end{array}
\]
which via Equation~(\ref{eq:KK_metric}), provides an explicit
expression for the two dimensional metric and the electromagnetic
vector potential
\begin{equation}
\label{eq:metric_2D_single_mode}
\begin{array}{lr}
\displaystyle
g^{\mu\nu}=\eta^{\mu\nu}+\sum_{\alpha=1}^2\left(-\eta^{\nu\alpha}\frac{\partial
u^\mu_0}{\partial x^\alpha}-\eta^{\mu\alpha}\frac{\partial
u^\nu_0}{\partial x^\alpha}+\sum_{\beta=1}^2\eta^{\alpha\beta}
\frac{\partial u^\mu_0}{\partial x^\alpha}\frac{\partial
u^\nu_0}{\partial x^\beta}\right),&\hspace{.1in}(\mu,\nu=1,2)
\end{array}
\end{equation}
and
\begin{equation}
\label{eq:A_single_mode}
\begin{array}{lr}
\displaystyle
 A^{\mu}=\sum_{\alpha=1}^2\left(-\eta^{\mu\alpha}\frac{\partial
u^3_0}{\partial x^\alpha}+\sum_{\beta=1}^2\eta^{\alpha\beta}
\frac{\partial u^\mu_0}{\partial x^\alpha}\frac{\partial
u^3_0}{\partial x^\beta}\right),&\hspace{.1in}(\mu=1,2)
\end{array}
\end{equation}
and where
 \begin{equation}
 \eta^{\mu\nu}\equiv \left(\begin{array}{ccc}
 -1 & 0 & 0\\
 0 & 1 & 0\\
 0 & 0 & 1
\end{array}
\right).
\end{equation}
Note that the
 2D metric in Equation~(\ref{eq:metric_2D_single_mode}) is no longer
 due to a simple coordinate change.  There is no coordinate
 transformation (or any other transformation) that involves only the
 coordinates $x'_1$ and $x'_2$ that will remove the Fourier
 transforms in this equation and globally create the form given in
 Equation~(\ref{eq:minkowski_metric}).

 Let us summarize what we have done in this section.
 We began with an equation for Dirac's equation in three-dimensions,
 Equation~(\ref{eq:dirac_curved_space}), which described a spin-$1/2$
 particle with zero mass and no interactions with the outside world.
 Using dimensional reduction we obtained a two-dimensional equation
 for a massive spin-$1/2$ particle in curved space, which interacts via the
 gravitational metric $g_{\mu\nu}$ and the
 electromagnetic potential $A_\mu$.

 We also began with a trivial flat space metric in three-dimensions,
 Equation~(\ref{eq:metric_tensor})
 and using dimensional reduction, derived a two-dimensional curved metric,
 Equation~(\ref{eq:metric_2D_single_mode}), and an electromagnetic
 vector potential, Equation~(\ref{eq:A_single_mode}).

 Equations~(\ref{eq:single_mode}), (\ref{eq:metric_2D_single_mode}) and
 (\ref{eq:A_single_mode}) represent the main result of this paper.
 They provide a quantum mechanical treatment of a combined system of
 the Dirac Equation,
 electromagnetism and gravity, albeit in a low dimensional setting.
 We note however that this whole procedure can be carried out in
 higher dimensions. The only part of the derivation that is lacking
 in higher dimensions
 is an explicit solution of the Dirac spinor in terms of fractional
 derivatives like those given in Equations~(\ref{sec:Cartan}).  If a
 higher dimensional version of this theory is constructed it would
 provide an alternate approach to current theories of quantum gravity.

\section{Conclusions}
We have taken a model of an elastic medium and derived an equation
of motion that has the same form as Dirac's equation in the presence
of electromagnetism and gravity.  We derived this equation by using
the formalism of Cartan to reduce the quadratic form of the wave
equation to the linear form of Dirac's equation. We showed that the
dimensional reduction technique from Kaluza Klein theory produces
not only the Einstein Field and Maxwell's equations but also induces
both mass and interaction terms into Dirac's Equation.   The
formalism demonstrates that a quantum mechanical treatment of the
 Einstein-Dirac-Maxwell equations can be derived from the
equations of motion of the Fourier modes of an elastic solid and
provides a new approach to theories of quantum gravity.
\appendix
\section{Quantization coefficients}
\label{app:quantization}
 In diagonalizing the Hamiltonian given in
Equation~(\ref{eq:hamiltonian_FT}) we first write the field
operators in terms of an intermediate set of ladder operators
\[
P_{n,q}=\frac{\imath}{\sqrt{2}}
\left(a^\dagger_{n,q}-a_{n,-q}\right)
\]
\[
u_{n,q}=\frac{1}{\sqrt{2}} \left(a_{n,q}+a^\dagger_{n,-q}\right)
\]
Now define the vectors
\begin{eqnarray*}
Q_q&=&\left(a_{1,q},a_{2,q},a_{3,q},a_{1,-q},a_{2,-q},a_{3,-q},a_{1,q}^\dagger
   ,a_{2,q}^\dagger,a_{3,q}^\dagger,a_{1,-q}^\dagger
   ,a_{2,-q}^\dagger ,a_{3,-q}^\dagger\right)\\
Q^\dagger_q&=&\left(a_{1,q}^\dagger ,a_{2,q}^\dagger
,a_{3,q}^\dagger
   ,a_{1,-q}^{\dagger },a_{2,-q}^\dagger,a_{3,-q}^\dagger
   ,a_{1,q},a_{2,q},a_{3,q},a_{1,-q},a_{2,-q},a_{3,-q}\right)
\end{eqnarray*}
This allows us to write $H_q=Q^\dagger_q A Q_q$ where the matrix $A$
is given by
\[
A=\left(
\begin{array}{ll}
T&S\\
S&T
\end{array}
\right)
\]
and the nonzero elements of the matrices $T$ and $S$ are
\begin{eqnarray*}
\small T_{1,1}&=& \frac{3}{16 (\lambda +2 \mu )}
\\
T_{1,5}&=& \frac{\mu  q_2}{4 (\lambda +2 \mu )}
\\
T_{1,6}&=& \frac{\mu  q_3}{4 (\lambda +2 \mu )}
\\
T_{2,2}&=& \frac{8 \left(2 q_2^2+q_3^2\right) \mu ^3+2 \mu +\lambda
\left(4 \left(4 q_2^2+q_3^2\right) \mu ^2+1\right)}{16 \mu  (\lambda
+2 \mu )}
\\
T_{2,3}&=& \frac{\mu  (3 \lambda +2 \mu ) q_2 q_3}{4 (\lambda +2 \mu
)}
\\
T_{2,4}&=& -\frac{\mu  q_2}{4 (\lambda +2 \mu )}
\\
T_{3,2}&=& \frac{\mu  (3 \lambda +2 \mu ) q_2 q_3}{4 (\lambda +2 \mu
)}
\\
T_{3,3}&=& \frac{8 \left(q_2^2+2 q_3^2\right) \mu ^3+2 \mu +\lambda
\left(4 \left(q_2^2+4 q_3^2\right) \mu ^2+1\right)}{16 \mu  (\lambda
+2 \mu )}
\\
T_{3,4}&=& -\frac{\mu  q_3}{4 (\lambda +2 \mu )}
\\
T_{4,2}&=& -\frac{\mu  q_2}{4 (\lambda +2 \mu )}
\\
T_{4,3}&=& -\frac{\mu  q_3}{4 (\lambda +2 \mu )}
\\
T_{4,4}&=& \frac{3}{16 (\lambda +2 \mu )}
\\
T_{5,1}&=& \frac{\mu  q_2}{4 (\lambda +2 \mu )}
\\
T_{5,5}&=& \frac{8 \left(2 q_2^2+q_3^2\right) \mu ^3+2 \mu +\lambda
\left(4 \left(4 q_2^2+q_3^2\right) \mu ^2+1\right)}{16 \mu  (\lambda
+2 \mu )}
\\
T_{5,6}&=& \frac{\mu  (3 \lambda +2 \mu ) q_2 q_3}{4 (\lambda +2 \mu
)}
\\
T_{6,1}&=& \frac{\mu  q_3}{4 (\lambda +2 \mu )}
\\
T_{6,5}&=& \frac{\mu  (3 \lambda +2 \mu ) q_2 q_3}{4 (\lambda +2 \mu
)}
\\
T_{6,6}&=& \frac{8 \left(q_2^2+2 q_3^2\right) \mu ^3+2 \mu +\lambda
\left(4 \left(q_2^2+4 q_3^2\right) \mu ^2+1\right)}{16 \mu  (\lambda
+2 \mu )}
\\
S_{1,8}&=& -\frac{(2 \lambda +2 \mu ) q_2}{8 (\lambda +2 \mu )}
\\
S_{1,9}&=& -\frac{(2 \lambda +2 \mu ) q_3}{8 (\lambda +2 \mu )}
\\
S_{1,10}&=& -\frac{3}{16 (\lambda +2 \mu )}
\\
S_{2,7}&=& -\frac{(2 \lambda +2 \mu ) q_2}{8 (\lambda +2 \mu )}
\\
S_{2,11}&=& \frac{2 \mu  \left(4 \mu ^2 \left(2
q_2^2+q_3^2\right)-1\right)+\lambda  \left(4 \mu ^2 \left(4
q_2^2+q_3^2\right)-1\right)}{16 \mu  (\lambda +2 \mu )}
\\
S_{2,12}&=& \frac{\mu  (3 \lambda +2 \mu ) q_2 q_3}{4 (\lambda +2
\mu )}
\\
S_{3,7}&=& -\frac{(2 \lambda +2 \mu ) q_3}{8 (\lambda +2 \mu )}
\\
S_{3,11}&=& \frac{\mu  (3 \lambda +2 \mu ) q_2 q_3}{4 (\lambda +2
\mu )}
\\
S_{3,12}&=& \frac{2 \mu  \left(4 \mu ^2 \left(q_2^2+2
q_3^2\right)-1\right)+\lambda  \left(4 \mu ^2 \left(q_2^2+4
q_3^2\right)-1\right)}{16 \mu  (\lambda +2 \mu )}
\\
S_{4,7}&=& -\frac{3}{16 (\lambda +2 \mu )}
\\
S_{4,11}&=& \frac{(2 \lambda +2 \mu ) q_2}{8 (\lambda +2 \mu )}
\\
S_{4,12}&=& \frac{(2 \lambda +2 \mu ) q_3}{8 (\lambda +2 \mu )}
\\
S_{5,8}&=& \frac{2 \mu  \left(4 \mu ^2 \left(2
q_2^2+q_3^2\right)-1\right)+\lambda  \left(4 \mu ^2 \left(4
q_2^2+q_3^2\right)-1\right)}{16 \mu  (\lambda +2 \mu )}
\\
S_{5,9}&=& \frac{\mu  (3 \lambda +2 \mu ) q_2 q_3}{4 (\lambda +2 \mu
)}
\\
S_{5,10}&=& \frac{(2 \lambda +2 \mu ) q_2}{8 (\lambda +2 \mu )}
\\
S_{6,8}&=& \frac{\mu  (3 \lambda +2 \mu ) q_2 q_3}{4 (\lambda +2 \mu
)}
\\
S_{6,9}&=& \frac{2 \mu  \left(4 \mu ^2 \left(q_2^2+2
q_3^2\right)-1\right)+\lambda  \left(4 \mu ^2 \left(q_2^2+4
q_3^2\right)-1\right)}{16 \mu  (\lambda +2 \mu )}
\\
S_{6,10}&=& \frac{(2 \lambda +2 \mu ) q_3}{8 (\lambda +2 \mu )}
\end{eqnarray*}
The methods outlined by Tikochinsky and
Tsallis\cite{ref:Tikochinsky,ref:Tsallis} can now be applied to the
matrix $A$. The final results allow us to define six creation
operators $b^\dagger_i$ and six annihilation operators $b_i$ that
satisfy
\[
\begin{array}{lcr}
[b_{i,q},b^\dagger_{j,q'}]=i\delta_{i,j}\delta_{q,q'}&\hspace{.1in}
[b_{i,q},b_{j,q'}]=0&\hspace{.1in}
[b^\dagger_{i,q},b^\dagger_{j,q'}]=0
\end{array}
\]
These operators are Eigenstates of the Hamiltonian with eigenvalues
\begin{eqnarray*}
E_{1,q}&=& \frac{1}{4} \sqrt{q_2^2+q_3^2}
\\
E_{2,q}&=& \frac{1}{4} \sqrt{q_2^2+q_3^2}
\\
E_{3,q}&=& \frac{1}{4} \sqrt{1-\frac{2 \sqrt{2} \sqrt{\mu  (\lambda
+2 \mu )^2 (2 \lambda +2 \mu )}}{(\lambda +2 \mu )^2}}
\sqrt{q_2^2+q_3^2}
\\
E_{4,q}&=& \frac{1}{4} \sqrt{1-\frac{2 \sqrt{2} \sqrt{\mu  (\lambda
+2 \mu )^2 (2 \lambda +2 \mu )}}{(\lambda +2 \mu )^2}}
\sqrt{q_2^2+q_3^2}
\\
E_{5,q}&=& \frac{1}{4} \sqrt{1+\frac{2 \sqrt{2} \sqrt{\mu  (\lambda
+2 \mu )^2 (2 \lambda +2 \mu )}}{(\lambda +2 \mu )^2}}
\sqrt{q_2^2+q_3^2}
\\
E_{6,q}&=& \frac{1}{4} \sqrt{1+\frac{2 \sqrt{2} \sqrt{\mu  (\lambda
+2 \mu )^2 (2 \lambda +2 \mu )}}{(\lambda +2 \mu )^2}}
\sqrt{q_2^2+q_3^2}
\\
\end{eqnarray*}
and the Hamiltonian has the diagonal form
\[
H_q=\sum_{i} E_{i,q}b^\dagger_{i,q} b_{i,q}
\]

The field displacement operators, $u_{ij}$ can be written in terms
of these ladder operators. Denote the vector of field and ladder
operators as
\begin{eqnarray*}
X_q&=&\left(P_{1,q},P_{2,q},P_{3,q},u_{1,q},u_{2,q},u_{3,q},P_{1,-q},P_{2,-q},P_{3,-q},u_{1,-q},u_{2,-q},u_{3,-q}\right)\\
B_q&=&\left(b_{1,q},b_{2,q},b_{3,q},b_{4,q},b_{5,q},b_{6,q},b^\dagger_{1,q},b^\dagger_{2,q},
b^\dagger_{3,q},b^\dagger_{4,q},b^\dagger_{5,q},b^\dagger_{6,q}\right).
\end{eqnarray*}
This allows us to write
\[
X_{i,q}=\sum_j c_{i,j}B_{i,q}
\]
where the coefficients $c_{i,j}$ are listed below.  In writing the
$c_{i,j}$ coefficients we have defined the following quantities
\begin{eqnarray*}
a&=&\sqrt{2} \sqrt{\mu  (\lambda +2 \mu )^2 (2 \lambda +2 \mu )}
\\
\lambda_2&=&\sqrt{1-\frac{2 \sqrt{2} \sqrt{\mu  (\lambda +2 \mu )^2
(2 \lambda +2 \ \mu )}}{(\lambda +2 \mu )^2}}
\\
\lambda_3&=&\sqrt{\frac{2 \sqrt{2} \sqrt{\mu  (\lambda +2 \mu )^2 (2
\lambda +2 \ \mu )}}{(\lambda +2 \mu )^2}+1}
\\
d_1&=&8 \mu  (2 \lambda +2 \mu ) \left(16 \text{Eq}^2 \mu ^3+16
\text{Eq}^2 \lambda  \mu ^2-2 \mu -\lambda \right) \\
&&\mbox{}\hspace{1in}+a \left(32 \text{Eq}^2
   \mu ^3+32 \text{Eq}^2 \lambda  \mu ^2-10 \mu -2 \lambda \right)
\\
d_2&=&8 \mu  (2 \lambda +2 \mu ) \left(16 \text{Eq}^2 \mu ^3+16
\text{Eq}^2 \lambda  \mu ^2-2 \mu -\lambda \right)\\
&&\mbox{}\hspace{1in}+a \left(-32 \text{Eq}^2
   \mu ^3+10 \mu +\lambda  \left(2-32 \text{Eq}^2 \mu ^2\right)\right)
\\
d_3&=&a \left(32 \text{Eq}^2 \mu ^3+32 \text{Eq}^2 \lambda  \mu
^2-10 \mu -2 \lambda \right)\\
&&\mbox{}\hspace{1in}-8 \mu  (2 \lambda +2 \mu ) \left(16
   \text{Eq}^2 \mu ^3+16 \text{Eq}^2 \lambda  \mu ^2-2 \mu -\lambda \right)
\\
d_4&=&12 a \mu +(\lambda +2 \mu ) \left(64 \text{Eq}^2 \mu ^4+96
\text{Eq}^2 \lambda  \mu ^3+4 \left(8 \text{Eq}^2 \lambda
^2+3\right) \mu
   ^2+18 \lambda  \mu -2 \lambda ^2\right)
\\d_5&=&(\lambda +2 \mu ) \left(64 \text{Eq}^2 \mu ^4+96 \text{Eq}^2 \lambda  \mu ^3+4 \left(8 \text{Eq}^2 \lambda ^2+3\right) \mu ^2+18
   \lambda  \mu -2 \lambda ^2\right)-12 a \mu
\\
 n_1&=&2 \sqrt{2}
\sqrt{\frac{\mu \left(q_2^2+q_3^2\right){}^{3/2}}{q_2^2
   \left(2 \mu  \omega _q-1\right){}^2}}\\
n_2&=&   2 \sqrt{2} \sqrt{\frac{\mu
\left(q_2^2+q_3^2\right){}^{3/2}}{q_2^2
   \left(2 \mu  \omega _q-1\right){}^2}}
   \\
{\textstyle n_3}&{\textstyle=}&  8 \sqrt{2}\biggl({\textstyle 8\mu
^2 (2 \lambda +2 \mu ) \omega _q^2 (2 (2 \mu  (\lambda +2 \mu )
\lambda _2^2 \omega _q^2+2 \mu  (3 \lambda +4 \mu )
   \omega _q^2 }
   \\
   &&\mbox{}{\textstyle +\lambda _2 (16 \omega _q^2 \mu ^3+2 \mu+\lambda  (16 \mu ^2 \omega _q^2+1)) \omega _q)
   a^2-(\lambda +2 \mu ) (4 \mu  (\lambda +2 \mu ) (2 \lambda +2 \mu ) \lambda _2^2 \omega _q^2}
   \\
   &&\mbox{}{\textstyle +4 \mu  (2 \lambda +2 \mu )
   (\lambda +10 \mu ) \omega _q^2+(\lambda +2 \mu ) \lambda _2 (2 \mu  (16 \mu ^2 \omega _q^2+5)+\lambda  (32
   \mu ^2 \omega _q^2+2)) \omega _q) a}
   \\
   &&\mbox{}{\textstyle +4 \mu  (\lambda +2 \mu )^2 (2 \lambda +2 \mu ) (2 \mu  (\lambda +2 \mu
   ) \lambda _2^2 \omega _q^2+2 \mu  (3 \lambda +4 \mu ) \omega _q^2}
   \\
   &&\mbox{}{\textstyle +\lambda _2 (16 \omega _q^2 \mu ^3+2 \mu +\lambda  (16
   \mu ^2 \omega _q^2+1)) \omega _q))}\biggr)^{1/2}
   /
   \biggl({\textstyle (\lambda +2 \mu ) q_3^2 (a (32 \omega _q^2 \mu ^3-10 \mu} \\
   &&\mbox{}{\textstyle +\lambda  (32 \mu ^2 \omega _q^2-2))-8 \mu  (2
   \lambda +2 \mu ) (16 \omega _q^2 \mu ^3-2 \mu +\lambda  (16 \mu ^2 \omega
   _q^2-1))){}^2}\biggr)^{1/2}
   \\
   n_4&=&8 \biggl( -\mu  (\lambda +2 \mu )^2 (8 (\lambda +2 \mu ) (2 \lambda +2
\mu ) (4 \lambda  \omega _q^2-3 \lambda _2 \omega _q)
   \mu ^2
   \\
   &&\mbox{}+a (\lambda _2 \omega _q ((32 \mu ^2 \omega _q^2+2) \lambda ^2+6 \mu  (16 \mu ^2 \omega
   _q^2-3) \lambda +4 \mu ^2 (16 \mu ^2 \omega _q^2-3))
   \\
   &&\mbox{}-8 (\lambda -6 \mu ) \mu  (2 \lambda +2 \mu ) \omega
   _q^2))\biggr)^{1/2}
   /
   \biggl( (12 a \mu +(\lambda +2 \mu ) ((32 \mu ^2 \omega _q^2-2) \lambda ^2
   \\
   &&\mbox{}+6 \mu  (16 \mu ^2 \omega _q^2+3)
   \lambda +4 \mu ^2 (16 \mu ^2 \omega
   _q^2+3)))^2\biggr)^{1/2}
   \\
   n_5&=&8\sqrt{2}\biggl(\mu ^2 (2 \lambda +2 \mu ) \omega _q^2 (2 (2 \mu  (\lambda +2 \mu ) \lambda _3^2 \omega _q^2+2 \mu  (3 \lambda +4 \mu )
   \omega _q^2\\
   &&\mbox{}+\lambda _3 (16 \omega _q^2 \mu ^3+2 \mu +\lambda  (16 \mu ^2 \omega _q^2+1)) \omega _q)
   a^2+(\lambda +2 \mu ) (4 \mu  (\lambda +2 \mu ) (2 \lambda +2 \mu ) \lambda _3^2 \omega _q^2\\
   &&\mbox{}+4 \mu  (2 \lambda +2 \mu )
   (\lambda +10 \mu ) \omega _q^2+(\lambda +2 \mu ) \lambda _3 (2 \mu  (16 \mu ^2 \omega _q^2+5)+\lambda  (32
   \mu ^2 \omega _q^2+2)) \omega _q) a\\
   &&\mbox{}+4 \mu  (\lambda +2 \mu )^2 (2 \lambda +2 \mu ) (2 \mu  (\lambda +2 \mu
   ) \lambda _3^2 \omega _q^2+2 \mu  (3 \lambda +4 \mu ) \omega _q^2+\lambda _3 (16 \omega _q^2 \mu ^3+2 \mu \\
   &&\mbox{}+\lambda  (16
   \mu ^2 \omega _q^2+1)) \omega _q))\biggr)^{1/2}
   /
   \biggl((\lambda +2 \mu ) q_3^2 (8 \mu  (2 \lambda +2 \mu ) (16 \omega _q^2 \mu ^3-2 \mu \\
   &&\mbox{}+\lambda  (16 \mu ^2 \omega
   _q^2-1))+a (32 \omega _q^2 \mu ^3-10 \mu +\lambda  (32 \mu ^2 \omega
   _q^2-2))){}^2\biggr)^{1/2}
   \\
   n_6&=&8\biggl(\mu  (\lambda +2 \mu )^2 (8 (\lambda +2 \mu ) (2 \lambda +2 \mu ) (3 \lambda _3 \omega _q-4 \lambda  \omega _q^2) \mu
   ^2\\
   &&\mbox{}+a (\lambda _3 \omega _q ((32 \mu ^2 \omega _q^2+2) \lambda ^2+6 \mu  (16 \mu ^2 \omega _q^2-3)
   \lambda +4 \mu ^2 (16 \mu ^2 \omega _q^2-3))\\
   &&\mbox{}-8 (\lambda -6 \mu ) \mu  (2 \lambda +2 \mu ) \omega
   _q^2))\biggr)^{1/2}
   /
   \biggl(((\lambda +2 \mu ) ((32 \mu ^2 \omega _q^2-2) \lambda ^2+6 \mu  (16 \mu ^2 \omega _q^2+3) \lambda \\
   &&\mbox{}+4
   \mu ^2 (16 \mu ^2 \omega _q^2+3))-12 a \mu
   ){}^2\biggr)^{1/2}
\end{eqnarray*}

Finally, the nonzero components of the matrix $c_{ij}$ are
\begin{eqnarray*}
{\textstyle c_{1,3}}&{\textstyle=}&{\textstyle \frac{i \sqrt{2}
\omega _q \left(a (2 \lambda -6 \mu ) (\lambda +2 \mu ) \lambda _2-4
\mu  (2 \lambda +2 \mu ) (4 \lambda  \mu  (\lambda +2 \mu )-a
(\lambda -6 \mu )) \omega _q\right)}{(\lambda +2 \mu ) d_2 n_3 q_3}
} \\
{\textstyle c_{1,5}}&{\textstyle=}&{\textstyle -\frac{i \sqrt{2}
\omega _q \left(a (2 \lambda -6 \mu ) (\lambda +2 \mu ) \lambda _3+4
\mu  (2 \lambda +2 \mu ) (a (\lambda -6 \mu )+4 \lambda  \mu
(\lambda +2 \mu )) \omega _q\right)}{(\lambda +2 \mu ) d_1 n_5 q_3}
} \\
{\textstyle c_{1,10}}&{\textstyle=}&{\textstyle \frac{i \sqrt{2}
\left(12 a \mu +(\lambda +2 \mu ) \left(-2 \lambda ^2+18 \mu \lambda
+12 \mu ^2+4 \mu  (2 a+(\lambda +2 \mu ) (2 \lambda +2 \mu ))
\lambda _2 \omega _q\right)\right)}{d_4 n_4}
} \\
{\textstyle c_{1,12}}&{\textstyle=}&{\textstyle \frac{i \sqrt{2}
\left((\lambda +2 \mu ) \left(-2 \lambda ^2+18 \mu  \lambda +12 \mu
^2+4 \mu  ((\lambda +2 \mu ) (2 \lambda +2 \mu )-2 a) \lambda _3
\omega _q\right)-12 a \mu \right)}{d_5 n_6}
} \\
{\textstyle c_{2,2}}&{\textstyle=}&{\textstyle -\frac{i \sqrt{2}
q_3}{n_2 q_2-2 \mu  n_2 q_2 \omega _q}
} \\
{\textstyle c_{2,4}}&{\textstyle=}&{\textstyle -\frac{i \sqrt{2}
(\lambda +2 \mu ) q_2 \left((\lambda +2 \mu ) \left(-2 \lambda
\lambda _2+6 \mu  \lambda _2+4 \mu  (2 \lambda +2 \mu ) \omega
_q\right)-8 a \mu  \omega _q\right)}{d_4 n_4 \omega _q}
} \\
{\textstyle c_{2,6}}&{\textstyle=}&{\textstyle -\frac{i \sqrt{2}
(\lambda +2 \mu ) q_2 \left(8 a \mu  \omega _q+(\lambda +2 \mu )
\left(-2 \lambda  \lambda _3+6 \mu  \lambda _3+4 \mu  (2 \lambda +2
\mu ) \omega _q\right)\right)}{d_5 n_6 \omega _q}
} \\
{\textstyle c_{2,7}}&{\textstyle=}&{\textstyle -\frac{i \sqrt{2}
q_3}{n_1 q_2-2 \mu  n_1 q_2 \omega _q}
} \\
{\textstyle c_{2,9}}&{\textstyle=}&{\textstyle \frac{i \sqrt{2} q_2
\left(a \left(2 \lambda +10 \mu +4 \mu  (2 \lambda +2 \mu ) \lambda
_2 \omega _q\right)-8 \mu  (\lambda +2 \mu ) (2 \lambda +2 \mu )
\left(2 \mu  \lambda _2 \omega _q+1\right)\right)}{d_2 n_3 q_3}
} \\
{\textstyle c_{2,11}}&{\textstyle=}&{\textstyle -\frac{i \sqrt{2}
q_2 \left(8 \mu  (\lambda +2 \mu ) (2 \lambda +2 \mu ) \left(2 \mu
\lambda _3 \omega _q+1\right)+a \left(2 \lambda +10 \mu +4 \mu  (2
\lambda +2 \mu ) \lambda _3 \omega _q\right)\right)}{d_1 n_5 q_3}
} \\
{\textstyle c_{3,2}}&{\textstyle=}&{\textstyle \frac{i
\sqrt{2}}{n_2-2 \mu  n_2 \omega _q}
} \\
{\textstyle c_{3,4}}&{\textstyle=}&{\textstyle -\frac{i \sqrt{2}
(\lambda +2 \mu ) q_3 \left((\lambda +2 \mu ) \left(-2 \lambda
\lambda _2+6 \mu  \lambda _2+4 \mu  (2 \lambda +2 \mu ) \omega
_q\right)-8 a \mu  \omega _q\right)}{d_4 n_4 \omega _q}
} \\
{\textstyle c_{3,6}}&{\textstyle=}&{\textstyle -\frac{i \sqrt{2}
(\lambda +2 \mu ) q_3 \left(8 a \mu  \omega _q+(\lambda +2 \mu )
\left(-2 \lambda  \lambda _3+6 \mu  \lambda _3+4 \mu  (2 \lambda +2
\mu ) \omega _q\right)\right)}{d_5 n_6 \omega _q}
} \\
{\textstyle c_{3,7}}&{\textstyle=}&{\textstyle \frac{i
\sqrt{2}}{n_1-2 \mu  n_1 \omega _q}
} \\
{\textstyle c_{3,9}}&{\textstyle=}&{\textstyle \frac{i \sqrt{2}
\left(a \left(2 \lambda +10 \mu +4 \mu  (2 \lambda +2 \mu ) \lambda
_2 \omega _q\right)-8 \mu  (\lambda +2 \mu ) (2 \lambda +2 \mu )
\left(2 \mu  \lambda _2 \omega _q+1\right)\right)}{d_2 n_3}
} \\
{\textstyle c_{3,11}}&{\textstyle=}&{\textstyle -\frac{i \sqrt{2}
\left(8 \mu  (\lambda +2 \mu ) (2 \lambda +2 \mu ) \left(2 \mu
\lambda _3 \omega _q+1\right)+a \left(2 \lambda +10 \mu +4 \mu  (2
\lambda +2 \mu ) \lambda _3 \omega _q\right)\right)}{d_1 n_5}
} \\
{\textstyle c_{4,4}}&{\textstyle=}&{\textstyle \frac{4 \sqrt{2} \mu
(\lambda +2 \mu ) \omega _q \left((\lambda +2 \mu ) (2 \lambda +2
\mu ) \left(4 \mu  \omega _q-\lambda _2\right)-2 a \lambda
_2\right)}{d_4 n_4}
} \\
{\textstyle c_{4,6}}&{\textstyle=}&{\textstyle \frac{4 \sqrt{2} \mu
(\lambda +2 \mu ) \omega _q \left(2 a \lambda _3+(\lambda +2 \mu )
(2 \lambda +2 \mu ) \left(4 \mu  \omega _q-\lambda
_3\right)\right)}{d_5 n_6}
} \\
{\textstyle c_{4,9}}&{\textstyle=}&{\textstyle -\frac{4 \sqrt{2} \mu
(2 \lambda +2 \mu ) \omega _q^2 \left(4 \mu  \lambda _2 \omega _q
a+a-4 \mu  (\lambda +2 \mu )\right)}{d_2 n_3 q_3}
} \\
{\textstyle c_{4,11}}&{\textstyle=}&{\textstyle \frac{4 \sqrt{2} \mu
(2 \lambda +2 \mu ) \omega _q^2 \left(4 \mu  \lambda _3 \omega _q
a+a+4 \mu  (\lambda +2 \mu )\right)}{d_1 n_5 q_3}
} \\
{\textstyle c_{5,1}}&{\textstyle=}&{\textstyle \frac{2 \sqrt{2} \mu
q_3 \omega _q}{n_1 q_2-2 \mu  n_1 q_2 \omega _q}
} \\
{\textstyle c_{5,3}}&{\textstyle=}&{\textstyle \frac{4 \sqrt{2} \mu
(2 \lambda +2 \mu ) q_2 \omega _q \left(a \left(\lambda _2+4 \mu
\omega _q\right)-4 \mu  \left((\lambda +2 \mu ) \lambda _2+4 \mu  (2
\lambda +2 \mu ) \omega _q\right)\right)}{d_3 n_3 q_3}
} \\
{\textstyle c_{5,5}}&{\textstyle=}&{\textstyle \frac{4 \sqrt{2} \mu
(2 \lambda +2 \mu ) q_2 \omega _q \left(a \left(\lambda _3+4 \mu
\omega _q\right)+4 \mu  \left((\lambda +2 \mu ) \lambda _3+4 \mu  (2
\lambda +2 \mu ) \omega _q\right)\right)}{d_1 n_5 q_3}
} \\
{\textstyle c_{5,8}}&{\textstyle=}&{\textstyle -\frac{2 \sqrt{2} \mu
q_3 \omega _q}{n_2 q_2-2 \mu  n_2 q_2 \omega _q}
} \\
{\textstyle c_{5,10}}&{\textstyle=}&{\textstyle -\frac{4 \sqrt{2}
\mu  q_2 \left(2 a \lambda +(\lambda +2 \mu ) (2 \lambda +2 \mu )
\left(-\lambda +6 \mu +4 \mu  (\lambda +2 \mu ) \lambda _2 \omega
_q\right)\right)}{d_4 n_4}
} \\
{\textstyle c_{5,12}}&{\textstyle=}&{\textstyle -\frac{4 \sqrt{2}
\mu  q_2 \left((\lambda +2 \mu ) (2 \lambda +2 \mu ) \left(-\lambda
+6 \mu +4 \mu  (\lambda +2 \mu ) \lambda _3 \omega _q\right)-2 a
\lambda \right)}{d_5 n_6}
} \\
{\textstyle c_{6,1}}&{\textstyle=}&{\textstyle -\frac{2 \sqrt{2} \mu
\omega _q}{n_1-2 \mu  n_1 \omega _q}
} \\
{\textstyle c_{6,3}}&{\textstyle=}&{\textstyle \frac{4 \sqrt{2} \mu
(2 \lambda +2 \mu ) \omega _q \left(a \left(\lambda _2+4 \mu  \omega
_q\right)-4 \mu  \left((\lambda +2 \mu ) \lambda _2+4 \mu  (2
\lambda +2 \mu ) \omega _q\right)\right)}{d_3 n_3}
} \\
{\textstyle c_{6,5}}&{\textstyle=}&{\textstyle \frac{4 \sqrt{2} \mu
(2 \lambda +2 \mu ) \omega _q \left(a \left(\lambda _3+4 \mu  \omega
_q\right)+4 \mu  \left((\lambda +2 \mu ) \lambda _3+4 \mu  (2
\lambda +2 \mu ) \omega _q\right)\right)}{d_1 n_5}
} \\
{\textstyle c_{6,8}}&{\textstyle=}&{\textstyle \frac{2 \sqrt{2} \mu
\omega _q}{n_2-2 \mu  n_2 \omega _q}
} \\
{\textstyle c_{6,10}}&{\textstyle=}&{\textstyle -\frac{4 \sqrt{2}
\mu  q_3 \left(2 a \lambda +(\lambda +2 \mu ) (2 \lambda +2 \mu )
\left(-\lambda +6 \mu +4 \mu  (\lambda +2 \mu ) \lambda _2 \omega
_q\right)\right)}{d_4 n_4}
} \\
{\textstyle c_{6,12}}&{\textstyle=}&{\textstyle -\frac{4 \sqrt{2}
\mu  q_3 \left((\lambda +2 \mu ) (2 \lambda +2 \mu ) \left(-\lambda
+6 \mu +4 \mu  (\lambda +2 \mu ) \lambda _3 \omega _q\right)-2 a
\lambda \right)}{d_5 n_6}
} \\
{\textstyle c_{7,4}}&{\textstyle=}&{\textstyle \frac{i \sqrt{2}
\left(12 a \mu +(\lambda +2 \mu ) \left(-2 \lambda ^2+18 \mu \lambda
+12 \mu ^2+4 \mu  (2 a+(\lambda +2 \mu ) (2 \lambda +2 \mu ))
\lambda _2 \omega _q\right)\right)}{d_4 n_4}
} \\
{\textstyle c_{7,6}}&{\textstyle=}&{\textstyle \frac{i \sqrt{2}
\left((\lambda +2 \mu ) \left(-2 \lambda ^2+18 \mu  \lambda +12 \mu
^2+4 \mu  ((\lambda +2 \mu ) (2 \lambda +2 \mu )-2 a) \lambda _3
\omega _q\right)-12 a \mu \right)}{d_5 n_6}
} \\
{\textstyle c_{7,9}}&{\textstyle=}&{\textstyle \frac{i \sqrt{2}
\omega _q \left(a (2 \lambda -6 \mu ) (\lambda +2 \mu ) \lambda _2-4
\mu  (2 \lambda +2 \mu ) (4 \lambda  \mu  (\lambda +2 \mu )-a
(\lambda -6 \mu )) \omega _q\right)}{(\lambda +2 \mu ) d_2 n_3 q_3}
} \\
{\textstyle c_{7,11}}&{\textstyle=}&{\textstyle -\frac{i \sqrt{2}
\omega _q \left(a (2 \lambda -6 \mu ) (\lambda +2 \mu ) \lambda _3+4
\mu  (2 \lambda +2 \mu ) (a (\lambda -6 \mu )+4 \lambda  \mu
(\lambda +2 \mu )) \omega _q\right)}{(\lambda +2 \mu ) d_1 n_5 q_3}
} \\
{\textstyle c_{8,1}}&{\textstyle=}&{\textstyle -\frac{i \sqrt{2}
q_3}{n_1 q_2-2 \mu  n_1 q_2 \omega _q}
} \\
{\textstyle c_{8,3}}&{\textstyle=}&{\textstyle \frac{i \sqrt{2} q_2
\left(a \left(2 \lambda +10 \mu +4 \mu  (2 \lambda +2 \mu ) \lambda
_2 \omega _q\right)-8 \mu  (\lambda +2 \mu ) (2 \lambda +2 \mu )
\left(2 \mu  \lambda _2 \omega _q+1\right)\right)}{d_2 n_3 q_3}
} \\
{\textstyle c_{8,5}}&{\textstyle=}&{\textstyle -\frac{i \sqrt{2} q_2
\left(8 \mu  (\lambda +2 \mu ) (2 \lambda +2 \mu ) \left(2 \mu
\lambda _3 \omega _q+1\right)+a \left(2 \lambda +10 \mu +4 \mu  (2
\lambda +2 \mu ) \lambda _3 \omega _q\right)\right)}{d_1 n_5 q_3}
} \\
{\textstyle c_{8,8}}&{\textstyle=}&{\textstyle -\frac{i \sqrt{2}
q_3}{n_2 q_2-2 \mu  n_2 q_2 \omega _q}
} \\
{\textstyle c_{8,10}}&{\textstyle=}&{\textstyle -\frac{i \sqrt{2}
(\lambda +2 \mu ) q_2 \left((\lambda +2 \mu ) \left(-2 \lambda
\lambda _2+6 \mu  \lambda _2+4 \mu  (2 \lambda +2 \mu ) \omega
_q\right)-8 a \mu  \omega _q\right)}{d_4 n_4 \omega _q}
} \\
{\textstyle c_{8,12}}&{\textstyle=}&{\textstyle -\frac{i \sqrt{2}
(\lambda +2 \mu ) q_2 \left(8 a \mu  \omega _q+(\lambda +2 \mu )
\left(-2 \lambda  \lambda _3+6 \mu  \lambda _3+4 \mu  (2 \lambda +2
\mu ) \omega _q\right)\right)}{d_5 n_6 \omega _q}
} \\
{\textstyle c_{9,1}}&{\textstyle=}&{\textstyle \frac{i
\sqrt{2}}{n_1-2 \mu  n_1 \omega _q}
} \\
{\textstyle c_{9,3}}&{\textstyle=}&{\textstyle \frac{i \sqrt{2}
\left(a \left(2 \lambda +10 \mu +4 \mu  (2 \lambda +2 \mu ) \lambda
_2 \omega _q\right)-8 \mu  (\lambda +2 \mu ) (2 \lambda +2 \mu )
\left(2 \mu  \lambda _2 \omega _q+1\right)\right)}{d_2 n_3}
} \\
{\textstyle c_{9,5}}&{\textstyle=}&{\textstyle -\frac{i \sqrt{2}
\left(8 \mu  (\lambda +2 \mu ) (2 \lambda +2 \mu ) \left(2 \mu
\lambda _3 \omega _q+1\right)+a \left(2 \lambda +10 \mu +4 \mu  (2
\lambda +2 \mu ) \lambda _3 \omega _q\right)\right)}{d_1 n_5}
} \\
{\textstyle c_{9,8}}&{\textstyle=}&{\textstyle \frac{i
\sqrt{2}}{n_2-2 \mu  n_2 \omega _q}
} \\
{\textstyle c_{9,10}}&{\textstyle=}&{\textstyle -\frac{i \sqrt{2}
(\lambda +2 \mu ) q_3 \left((\lambda +2 \mu ) \left(-2 \lambda
\lambda _2+6 \mu  \lambda _2+4 \mu  (2 \lambda +2 \mu ) \omega
_q\right)-8 a \mu  \omega _q\right)}{d_4 n_4 \omega _q}
} \\
{\textstyle c_{9,12}}&{\textstyle=}&{\textstyle -\frac{i \sqrt{2}
(\lambda +2 \mu ) q_3 \left(8 a \mu  \omega _q+(\lambda +2 \mu )
\left(-2 \lambda  \lambda _3+6 \mu  \lambda _3+4 \mu  (2 \lambda +2
\mu ) \omega _q\right)\right)}{d_5 n_6 \omega _q}
} \\
{\textstyle c_{10,3}}&{\textstyle=}&{\textstyle \frac{4 \sqrt{2} \mu
(2 \lambda +2 \mu ) \omega _q^2 \left(4 \mu  \lambda _2 \omega _q
a+a-4 \mu  (\lambda +2 \mu )\right)}{d_2 n_3 q_3}
} \\
{\textstyle c_{10,5}}&{\textstyle=}&{\textstyle -\frac{4 \sqrt{2}
\mu  (2 \lambda +2 \mu ) \omega _q^2 \left(4 \mu  \lambda _3 \omega
_q a+a+4 \mu  (\lambda +2 \mu )\right)}{d_1 n_5 q_3}
} \\
{\textstyle c_{10,10}}&{\textstyle=}&{\textstyle -\frac{4 \sqrt{2}
\mu  (\lambda +2 \mu ) \omega _q \left((\lambda +2 \mu ) (2 \lambda
+2 \mu ) \left(4 \mu  \omega _q-\lambda _2\right)-2 a \lambda
_2\right)}{d_4 n_4}
} \\
{\textstyle c_{10,12}}&{\textstyle=}&{\textstyle -\frac{4 \sqrt{2}
\mu  (\lambda +2 \mu ) \omega _q \left(2 a \lambda _3+(\lambda +2
\mu ) (2 \lambda +2 \mu ) \left(4 \mu  \omega _q-\lambda
_3\right)\right)}{d_5 n_6}
} \\
{\textstyle c_{11,2}}&{\textstyle=}&{\textstyle \frac{2 \sqrt{2} \mu
q_3 \omega _q}{n_2 q_2-2 \mu  n_2 q_2 \omega _q}
} \\
{\textstyle c_{11,4}}&{\textstyle=}&{\textstyle \frac{4 \sqrt{2} \mu
q_2 \left(2 a \lambda +(\lambda +2 \mu ) (2 \lambda +2 \mu )
\left(-\lambda +6 \mu +4 \mu  (\lambda +2 \mu ) \lambda _2 \omega
_q\right)\right)}{d_4 n_4}
} \\
{\textstyle c_{11,6}}&{\textstyle=}&{\textstyle \frac{4 \sqrt{2} \mu
q_2 \left((\lambda +2 \mu ) (2 \lambda +2 \mu ) \left(-\lambda +6
\mu +4 \mu  (\lambda +2 \mu ) \lambda _3 \omega _q\right)-2 a
\lambda \right)}{d_5 n_6}
} \\
{\textstyle c_{11,7}}&{\textstyle=}&{\textstyle -\frac{2 \sqrt{2}
\mu  q_3 \omega _q}{n_1 q_2-2 \mu  n_1 q_2 \omega _q}
} \\
{\textstyle c_{11,9}}&{\textstyle=}&{\textstyle -\frac{4 \sqrt{2}
\mu  (2 \lambda +2 \mu ) q_2 \omega _q \left(a \left(\lambda _2+4
\mu  \omega _q\right)-4 \mu  \left((\lambda +2 \mu ) \lambda _2+4
\mu  (2 \lambda +2 \mu ) \omega _q\right)\right)}{d_3 n_3 q_3}
} \\
{\textstyle c_{11,11}}&{\textstyle=}&{\textstyle -\frac{4 \sqrt{2}
\mu  (2 \lambda +2 \mu ) q_2 \omega _q \left(a \left(\lambda _3+4
\mu  \omega _q\right)+4 \mu  \left((\lambda +2 \mu ) \lambda _3+4
\mu  (2 \lambda +2 \mu ) \omega _q\right)\right)}{d_1 n_5 q_3}
} \\
{\textstyle c_{12,2}}&{\textstyle=}&{\textstyle -\frac{2 \sqrt{2}
\mu  \omega _q}{n_2-2 \mu  n_2 \omega _q}
} \\
{\textstyle c_{12,4}}&{\textstyle=}&{\textstyle \frac{4 \sqrt{2} \mu
q_3 \left(2 a \lambda +(\lambda +2 \mu ) (2 \lambda +2 \mu )
\left(-\lambda +6 \mu +4 \mu  (\lambda +2 \mu ) \lambda _2 \omega
_q\right)\right)}{d_4 n_4}
} \\
{\textstyle c_{12,6}}&{\textstyle=}&{\textstyle \frac{4 \sqrt{2} \mu
q_3 \left((\lambda +2 \mu ) (2 \lambda +2 \mu ) \left(-\lambda +6
\mu +4 \mu  (\lambda +2 \mu ) \lambda _3 \omega _q\right)-2 a
\lambda \right)}{d_5 n_6}
} \\
{\textstyle c_{12,7}}&{\textstyle=}&{\textstyle \frac{2 \sqrt{2} \mu
\omega _q}{n_1-2 \mu  n_1 \omega _q}
} \\
{\textstyle c_{12,9}}&{\textstyle=}&{\textstyle -\frac{4 \sqrt{2}
\mu  (2 \lambda +2 \mu ) \omega _q \left(a \left(\lambda _2+4 \mu
\omega _q\right)-4 \mu  \left((\lambda +2 \mu ) \lambda _2+4 \mu  (2
\lambda +2 \mu ) \omega _q\right)\right)}{d_3 n_3}
} \\
{\textstyle c_{12,11}}&{\textstyle=}&{\textstyle -\frac{4 \sqrt{2}
\mu  (2 \lambda +2 \mu ) \omega _q \left(a \left(\lambda _3+4 \mu
\omega _q\right)+4 \mu  \left((\lambda +2 \mu ) \lambda _3+4 \mu  (2
\lambda +2 \mu ) \omega _q\right)\right)}{d_1 n_5}
} \\
\end{eqnarray*}
\section{Auxiliary Equation}
\label{app:Auxiliary_Equation} We wish to show that
\[
\frac{\partial {\tilde{\gamma}^\mu}}{\partial
x^\nu}+\tilde{\gamma}^\beta\Gamma^{\prime\mu}_{\beta\nu}-
\Gamma_\nu\tilde{\gamma}^\mu+\tilde{\gamma}^\mu\Gamma_\nu=0
\]
We first consider the equation
\[
\frac{\partial \gamma^\mu}{\partial x^\nu}=0
\]
true in the unprimed coordinate system.  But since the unprimed
coordinate system is Euclidean space, the Christoffel symbols are
identically zero. This allows us to write
\[
\frac{\partial \gamma^\mu}{\partial
x^\nu}+\gamma^\beta\Gamma^\mu_{\beta\nu}=0
\]
Since this is a tensor equation true in all frames, in the primed
coordinate system we can immediately write
\[
\partial'_\nu
\gamma^{\prime\mu}+\gamma^{\prime\beta}\Gamma^{\prime\mu}_{\beta\nu}=0
\]
Using $\gamma^{\prime\mu}=S\tilde{\gamma}^\mu S^{-1}$, we have

\[
 \partial'_\nu (S\tilde{\gamma}^\mu
S^{-1})+(S\tilde{\gamma}^\beta S^{-1})\Gamma'^\mu_{\beta\nu} =0
\]
or
\[(\partial'_\nu
S)\tilde{\gamma}^\mu S^{-1}+ S(\partial'_\nu \tilde{\gamma}^\mu)
S^{-1}+ S \tilde{\gamma}^\mu (\partial'_\nu
S^{-1})+(S\tilde{\gamma}^\beta S^{-1})\Gamma'^\mu_{\beta\nu}=0.
\]

Multiplying by $S^{-1}$ on the left and $S$ on the right yields
\[
S^{-1}(\partial'_\nu S)\tilde{\gamma}^\mu + (\partial'_\nu
\tilde{\gamma}^\mu) + \tilde{\gamma}^\mu (\partial'_\nu
S^{-1})S+\tilde{\gamma}^\beta \Gamma'^\mu_{\beta\nu} =0
\]

Finally, using $\Gamma_\nu=(\partial'_\nu S^{-1})S$ and again noting
that $\partial'_\nu S^{-1}S=-S^{-1}\partial'_\nu S$ we have,
\begin{equation}
\label{eq:aux_3D}
\tilde{\gamma}^\mu\Gamma_\nu-\Gamma_\nu\tilde{\gamma}^\mu
+\left(\partial'_\nu \tilde{\gamma}^\mu  +\tilde{\gamma}^\beta
\Gamma'^\mu_{\beta\nu}\right)=0
\end{equation}

\section{Decomposition of $\tilde{\gamma}^{\prime 3}$}
\label{app:gamma_perp}
 In decomposing the matrix
$\tilde{\gamma}^{\prime 3}$ we seek a matrix $\gamma_\bot$ that
satisfies
\begin{equation}
\label{eq:perpindicular_gamma}
\begin{array}{ccc}
\left\{\tilde{\gamma}^1,\gamma_\bot\right\}=0,&\hspace{.1in}
\left\{\tilde{\gamma}^2,\gamma_\bot\right\}=0,&
\hspace{.1in}\gamma_\bot^2=1
\end{array}
\end{equation}
where from Equations~(\ref{eq:tilde_gamma}) and
(\ref{eq:modified_gamma_matrices}) we can write
\begin{eqnarray*}
\tilde{\gamma}^1&=&S^{-1}\biggl((1-u^1_1)\gamma^1+u^1_2\gamma^2+u^1_3
\gamma^3\biggr)S
\\
 \tilde{\gamma}^2&=&S^{-1}\biggl(u^2_1
\gamma^1+(1-u^2_2)\gamma^2+u^2_3\gamma^3\biggr)S
\end{eqnarray*}
We can treat the matrices $\gamma^\mu$ as vectors and use the cross
product formula to compute an orthogonal vector.  In other words we
can write
\[
\gamma_\bot=\tilde{\gamma}^1\times\tilde{\gamma}^2
\]
using the rules
\[
\begin{array}{lcr}
\gamma^1\times\gamma^2=\gamma^3&\hspace{.1in}
\gamma^2\times\gamma^3=\gamma^1&\hspace{.1in}
\gamma^1\times\gamma^3=-\gamma^2
\end{array}.
\]
This gives
\begin{eqnarray*}
v_\bot&=&S^{-1}\biggl([u^1_2
u^2_3-u^1_3(1-u^2_2)]\gamma^1+[u^1_3u^2_1-(1-u^1_1)u^2_3]\gamma^2\\
&&\mbox{}\hspace{2in}+[(1-u^1_1)(1-u^2_2)-u^1_2u^2_1]\gamma^3\biggr)S
\end{eqnarray*}
with
\[
\gamma_\bot\equiv \frac{v_\bot}{|v_\bot|}
\]
and $|v_\bot|=\sqrt{v_\bot^2}$. It can be verified directly that
$\gamma_\bot$ satisfies Equation~(\ref{eq:perpindicular_gamma}).
\section{Fourier components}
\label{app:Fcomponents} We wish to relate the Fourier components of
the field displacements $u_i$, when expanded in the basis $e^{\imath
q' x'}$ to those expanded in the basis $e^{\imath q x}$.  The two
coordinate systems are related by $u_\mu=x_\mu-x'_\mu$ where the
field displacements $u_i$ are assumed to be small. We write
\begin{eqnarray*}
u_\mu&=&\sum_q u_{\mu q} e^{\imath qx}\\
&=&\sum_q u_{\mu q} e^{\imath q\cdot(x'+u)}\\
&\approx&\sum_q u_{\mu q} e^{\imath qx'}(1+\imath q\cdot u)\\
&=&\sum_q u_{\mu q} e^{\imath qx'}(1+\imath \sum_k q\cdot
u_{k}e^{\imath k\cdot x})\\
&\approx&\sum_q u_{\mu q} e^{\imath qx'}(1+\imath \sum_k q\cdot
u_{k}e^{\imath k\cdot x'})\\
&=&\sum_{qk}u_{\mu q}(\delta_{k,0}+\imath q\cdot u_k)e^{\imath
(q+k)\cdot x'}
\end{eqnarray*}
correct to second order in $u_{\mu,q}$.

Setting $q+k=q'$ we have
\[
u_\mu=\sum_{q'k}u_{\mu, q'-k}(\delta_{k,0}+\imath (q'-k)\cdot
u_k)e^{\imath q'\cdot x'}
\] so that the components of the field displacements in the primed
frame are
\[
u_{m'}=\sum_k u_{m'-k}(\delta_{k,0}+\imath(m'-k)u_k)
\]
We can now relate the energy eigenstates which are expressed in the
unprimed frame to the Fourier components in the primed frame.  In a
low energy theory in which only $u_0$ is present, to second order in
$u_{i}$, the only Fourier component in the primed frame that is
nonzero is $u_{m'=0}$. If there is enough energy to excite the
fields $u_0$ and $u_\pm1$. Then the only nonzero modes in the primed
frame are $u_{m'=0}$ and $u_{m'=\pm1}$ and so on.

\end{document}